# Origin of $Fe^{3+}$ in Fe-containing, Al-free Mantle Silicate Perovskite

Shenzhen Xu[1], Sang-Heon Shim[3], Dane Morgan[1,2]

[1]*Materials Science Program*

[2]*Department of Materials Science and Engineering*

*University of Wisconsin – Madison, Madison, WI*

[3]*School of Earth and Space Exploration*

*Arizona State University, Tempe, AZ*

Corresponding author: Telephone: +1-608-265-5879; Fax: +1-608-262-8353; Email: ddmorgan@wisc.edu

**Abstract**

We have studied the ferrous ($Fe^{2+}$) and ferric ($Fe^{3+}$) iron concentrations in Al-free Fe containing Mg-silicate perovskite (Mg-Pv) at pressure (*P*), temperature (*T*), and oxygen fugacity ($fO_2$) conditions related to the lower mantle using a thermodynamic model based on ab-initio calculations. We consider the oxidation reaction and the charge disproportionation reaction, both of which can produce $Fe^{3+}$ in Mg-Pv. The model shows qualitatively good agreement with available experimental data on $Fe^{3+}/\Sigma Fe$ ($\Sigma Fe$ = total Fe in system), spin transitions, and equations of state. We predict that under lower-mantle conditions $Fe^{3+}/\Sigma Fe$ determined by the charge disproportionation is estimated to be 0.01-0.07 in Al-free Mg-Pv, suggesting that low Al Mg-Pv in the uppermost pyrolitic mantle (where majoritic garnet contains most of the Al) and in the harzburgitic heterogeneities throughout the lower mantle contains very little $Fe^{3+}$. We find that the volume reduction by the spin transition of the B-site $Fe^{3+}$ leads to a minimum $Fe^{3+}/\Sigma Fe$ in Mg-Pv at mid-mantle pressures. The model shows that configurational entropy is a key driving force to create $Fe^{3+}$ and therefore $Fe^{3+}$ content is highly temperature sensitive. The temperature sensitivity may lead to a maximum $Fe^{3+}/\Sigma Fe$ in Mg-Pv in warm regions at the core-mantle boundary region, such as Large Low Shear Velocity Provinces (LLSVPs), potentially altering the physical (e.g., bulk modulus) and transport (e.g., thermal and electrical conductivities) properties of the heterogeneities.

## 1. Introduction

Mg-silicate perovskite (Mg-Pv, bridgmanite) is the most abundant mineral phase in the lower mantle of the Earth. More than 70 percent of the lower mantle may be made up of Mg-Pv by volume and iron is the most abundant transition metal element in the lower mantle (Irifune et al., 2010). Iron is known to exist in Mg-Pv as both $Fe^{2+}$ and $Fe^{3+}$ (McCammon, 1997; Frost et al., 2004). The presence of $Fe^{3+}$ can influence a number of physical and chemical properties of Mg-Pv. For example, the high spin to low spin transition of $Fe^{3+}$ in the B-site of Mg-Pv may cause an anomalous behavior of the bulk modulus affecting Mg-Pv mechanical behavior (Hsu et al., 2011). This spin transition of $Fe^{3+}$ may also lead to a volume collapse and increase the density (Catalli et al., 2011, 2010; Hsu et al., 2011). Moreover an increase in the $Fe^{3+}$ concentration has been shown to increase the electrical conductivity (McCammon, 1997) and radiative thermal conductivity (Goncharov et al., 2008) of Mg-Pv. Many experiments have shown that even if the starting materials do not contain any $Fe^{3+}$, Mg-Pv synthesized at lower mantle pressures-temperature ($P$-$T$) conditions contains $Fe^{3+}$ (Frost and Langenhorst, 2002; Sinmyo et al., 2008; Grocholski et al., 2009). The $Fe^{3+}$ concentration in Mg-Pv ($[Fe^{3+}]/([Fe^{2+}] + [Fe^{3+}]) = Fe^{3+}/\Sigma Fe$, where $[X]$ is the number of moles of $X$ ranges from 0.1 to 0.6. Furthermore, the amount of $Fe^{3+}$ is very different between Al-free and Al-bearing Mg-Pv (Frost et al., 2004; McCammon et al., 2004; Nakajima et al., 2012), with Al bearing Mg-Pv containing much more $Fe^{3+}$.

Two mechanisms have been considered to dominate the formation of $Fe^{3+}$ in Mg-Pv (Frost et al., 2004; Nakajima et al., 2012; Zhang and Oganov, 2006): (a) the oxidation reaction, where $Fe^{2+}$ is oxidized to $Fe^{3+}$ under elevated oxygen fugacity ($fO_2$), and (b) the charge disproportionation (chg. disp.) reaction, where Fe spontaneously disproportionates through $3Fe^{2+} => 2Fe^{3+} + Fe^0$. The oxidation reactions in our work are written in terms of a reaction with $O_2$ gas with a specific $fO_2$. This approach avoids the complexity of tracking all the possible reduction reactions that might release oxygen from the oxide phases as we simply represent their combined contribution to the environment by $O_2$ gas with a given $fO_2$. This approach also has the advantage of providing a common reference for the oxidative strength of the environment that can be used to consider different conditions. However, we note that under lower mantle conditions it is expected that essentially all the oxygen is stored in solid phases and any significant reactions incorporating oxygen will take place through reduction and oxygen release of other oxide phases. As we will show below, under these conditions no excess oxygen is available for oxidation of $Fe^{2+}$, the $fO_2$ is very low, and only chg. disp. can change the $Fe^{3+}/\Sigma Fe$ ratio.

Because $Fe^{3+}$ content may depend on $fO_2$, it is important to control $fO_2$ in experiments to the extent possible. In some cases, metallic iron powder was mixed with the starting material to ensure reducing conditions (Frost et al., 2004; Grocholski et al., 2009), although the exact $fO_2$ is still not known. In most diamond-anvil cell experiments, $fO_2$ has not been carefully regulated or characterized. Because of the limited sample chamber space in a multi-anvil press, the capsule materials have been used to control the $fO_2$ in the chamber, such as Re and diamond capsules (Lautherbach et al., 2000; Frost et al., 2004). In this work, we use the $fO_2$ of the capsule materials to approximate the $fO_2$ in the sample chamber (Nakajima et al., 2012) and the details of the determination of $fO_2$ at different $T$-$P$ are shown in Supplemental information (SI) section 1.

The chg. disp. reaction has been studied in both experiments (Frost et al., 2004; McCammon, 1997) and simulations (Zhang and Oganov, 2006). Whereas the experiments suggest $Fe^{3+}/\Sigma Fe = 0\sim0.3$ in Al-free Mg-Pv, the simulations imply spontaneous oxidation of all Fe to $Fe^{3+}$ (and therefore $Fe^{3+}/\Sigma Fe \sim 1$) in Al-free Mg-Pv. Furthermore, a number of uncertainties remain about $Fe^{3+}$ in Mg-Pv, including the extent to which each of the two mechanisms (oxidation reaction and chg. disp. reaction) contributes more to its formation, the balance of enthalpic and entropic driving forces governing $Fe^{3+}$ formation (which will control the temperature dependence of the $Fe^{3+}$ concentration), and what $Fe^{3+}$ - $fO_2$ relationship exists at lower mantle $P$-$T$ conditions, which are difficult to achieve in laboratory experiments. In this study we develop an ab-initio based thermodynamic model to understand and quantify the mechanisms that produce $Fe^{3+}$ at high $P$-$T$. In our detailed models for generating $Fe^{3+}$ in Mg-Pv we do not include any mechanisms involving creation of oxygen vacancies, although we do consider the energetics of relevant reactions in section 2.1.1. There is still some debate about whether oxygen vacancy substitution plays a significant role when $Fe^{3+}$ enters into the Mg-Pv and is evidence from previous experimental work (McCammon, 1998) suggesting that oxygen vacancy substitution may be the dominant mechanism in Al-free Mg-Pv. However, some previous experimental and simulation studies (Lauterbach et al., 2000; Frost et al., 2002; Brodholt et al., 2000; Zhang and Oganov, 2006) as well as our own calculations (see Eq. 2 in section 2.1.1) indicate that the charge-coupled substitution considered in this work is much more favorable than oxygen vacancy mechanisms under the lower mantle relevant conditions. In particular, in Zhang and Oganov's (Zhang and Oganov, 2006) work, they explicitly showed that oxygen vacancy substitution is unfavorable compared with charge-coupled substitution in Al-free Mg-Pv, which is consistent with our calculation in section 2.1.1. Given that we do not predict the oxygen vacancy substitution mechanism to be active we do not include the mechanism in our detailed thermodynamic models.

We focus on Al-free Mg-Pv in order to avoid the large increase in complexity associated with the presence of Al. This simpler model can help us understand the origin of $Fe^{3+}$ and its dependence on different factors more easily than one containing Al. In addition, Al-free Mg-Pv is of interest for comparison to multiple Al free experiments as well as the upper part of the lower mantle and some subducting slabs, which are expected to have low Al concentration.

We will answer the following questions. (1) what are the dominant processes and thermodynamics driving forces that produce $Fe^{3+}$ at experimental and lower mantle conditions, (2) what is the effect of spin transition, pressures and temperatures on $Fe^{3+}$ concentration, and (3) what is the dependence of $Fe^{3+}/\Sigma Fe$ on $fO_2$ in Al-free Mg-Pv. The understanding of the thermodynamics of the Al-free Mg-Pv obtained from this work will guide researchers in building a more complete Al-bearing model in the future.

## 2. Modeling
### 2.1 Thermodynamic model

### 2.1.1 Oxidation model

The oxidation reaction occurs at a certain oxidation potential, which is represented by oxygen gas at a specific $fO_2$. As noted in section 1, the use of $O_2$ gas to describe $\mu(O_2)$ is consistent with the fact that the oxygen in the lower mantle is stored in a solid form. We assume the oxidation reaction occurs via the charge-coupled substitution (Lauterbach et al., 2000; Frost et al., 2002; Zhang and Oganov, 2006) mechanism, where MgO is incorporated and $Fe^{2+}$ is oxidized and substitutes on the B-site (see below for discussion of another other possible vacancy mediated oxidation reaction). The equation describing this oxidation reaction under conditions of excess MgO can be written:

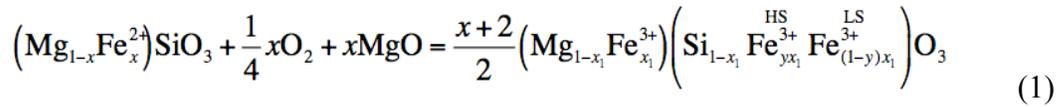

$$\left(Mg_{1-x}Fe^{2+}_x\right)SiO_3 + \frac{1}{4}xO_2 + xMgO = \frac{x+2}{2}\left(Mg_{1-x_1}Fe^{3+}_{x_1}\right)\left(Si_{1-x_1}\ Fe^{3+,HS}_{yx_1}\ Fe^{3+,LS}_{(1-y)x_1}\right)O_3 \qquad (1)$$

where $x_1 = x / (x+2)$, $Fe^{HS}$ and $Fe^{LS}$ refer to the high-spin and low-spin Fe respectively. The first term in the left hand side is ferrous Mg-Pv and $x$ is the fraction of Fe in the A site. The right hand side material is ferric Mg-Pv. The total Fe content in Mg-Pv is approximately 0.1~0.2 in previous experimental and simulation studies and also in the lower mantle (Auzende et al., 2008; Frost et al., 2004; Hsu et al., 2011; Irifune et al., 2010; Mao et al., 2011b; Grocholski et al., 2009). In the following calculations we set $x = 0.125$. In most cases one is interested in Mg-Pv coexisting with ferropericlase (Mg,Fe)O and there will be Fe partitioning between the phases. This partitioning can change the content of Fe in the Mg-Pv but such changes turn out to have a minor effect on the $Fe^{3+}/\Sigma Fe$ ratio within the Mg-Pv in the Al-free case. The impact of the Fe partitioning on the $Fe^{3+}/\Sigma Fe$ in is discussed in SI section 4.

Some previous experimental and simulation studies showed that in the lower-mantle pressure range up to 130 GPa, there is a HS to LS transition for Fe on only the B-site $Fe^{3+}$ (Bengtson et al., 2008; Catalli et al., 2010; Hsu et al., 2012, 2011; Lin et al., 2012). In Eq. (1), the parameter $y$ gives the fraction of $Fe^{3+}$ that is high spin in the B site.

In this work we do not include the production of $Fe^{3+}$ by oxygen vacancy substitution in the models and we assume all the $Fe^{3+}$ enters into Mg-Pv by charge-coupled substitution, as described in Eq. (1). Previous ab initio study of the $Al^{3+}$ incorporation into the Mg-Pv suggested that compared with charge-coupled substitution, oxygen vacancy substitution is unfavorable above 30 GPa (Brodholt, 2000), and here we demonstrate that this result holds for the ab initio methods used in this paper as well. Following the approach of Zhang and Oganov (2006), in order to evaluate whether oxygen vacancy substitution is favorable, we can study the following reaction:

$$Mg_{24}(Si_{22}Fe_2)O_{71} = (Mg_{22}Fe)(Si_{22}Fe)O_{69} + 2MgO \qquad (2)$$

It represents the reaction from an oxygen vacancy substitution to a charge coupled substitution. Our calculated reaction enthalpy $\Delta H$ is -6.23 eV at $P = 40$ GPa, which means the charge coupled substitution is much more stable than the oxygen vacancy substitution at lower-mantle conditions.

In order to model the free energies of reactions, the enthalpies are determined from ab initio calculations and the entropies are determined from analytical models. Configuration entropy is assumed to be ideal for all mixed occupation sublattices. The magnetic-electronic entropy is modeled based on totally disordered spins. It is assumed that the vibrational free energy of solid phase atoms cancels between reactants and products in Eq. (1) and the vibrational free energy difference between the $O_2$ gas and the $O^{2-}$ ions of the solid is treated by applying an empirical thermodynamic free energy model for $O_2$ gas and a simple Einstein model (with Einstein temperature 500 K) for solid $O^{2-}$ ions (Lee and Morgan, 2012). The empirical free energy expression for the oxygen atom (gas phase) is: $\frac{1}{2}(H_{O_2}^{NIST} - TS_{O_2}^{NIST} + kT \ln fO_2)$. $(H_{O_2}^{NIST} - TS_{O_2}^{NIST})$ is the empirical free energy of $O_2$ at $P^0(O_2) = 1$ bar, referenced to $H_{O_2}^{NIST}$ ($T = 298.15$ K). $kT \ln fO_2$ is the contribution to the oxygen free energy due to the change of $P(O_2)$ from the reference state of 1 atm. For the solid $O^{2-}$ ions, the free energy based on a simple Einstein model is: $(G_{vib}(O^{2-}_{solid}) - H^0_{vib}(O^{2-}_{solid}))$ with Einstein temperature = 500 K, which is referenced to $H^0_{vib}(O^{2-}_{solid})$ ($T = 298.15$ K). The exact expressions and necessary parameters of all the terms above can be found in Lee and Morgan (2012). The energy correction term of $O_2$ due to the over-binding from DFT is very small in HSE06 (Chevrier et al., 2010) and can be ignored. The $fO_2$ values at high P-T are extrapolated from experimental results (Campbell et al., 2007). Please see the details in the SI section 1.

To simplify the modeling, we write the whole system, both reactants and products in Eq. (1), in a compact form as

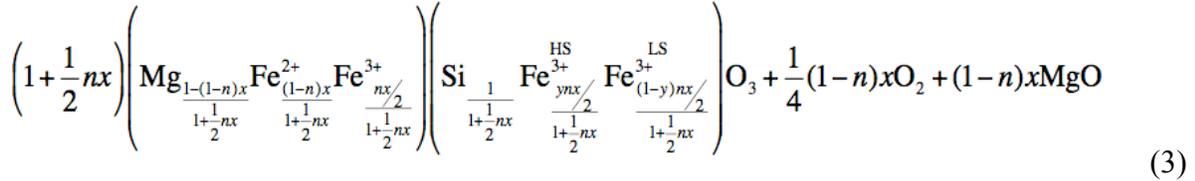

$$\left(1+\frac{1}{2}nx\right)\left(Mg_{\frac{1-(1-n)x}{1+\frac{1}{2}nx}}Fe^{2+}_{\frac{(1-n)x}{1+\frac{1}{2}nx}}Fe^{3+}_{\frac{nx/2}{1+\frac{1}{2}nx}}\right)\left(Si_{\frac{1}{1+\frac{1}{2}nx}}\overset{HS}{Fe^{3+}_{\frac{ynx/2}{1+\frac{1}{2}nx}}}\overset{LS}{Fe^{3+}_{\frac{(1-y)nx/2}{1+\frac{1}{2}nx}}}\right)O_3 + \frac{1}{4}(1-n)xO_2 + (1-n)xMgO \quad (3)$$

The parameter $n$ gives the fraction of ferrous Mg-Pv transformed to ferric Mg-Pv. Note that it is assumed that the reactants and products can be mixed into the same Mg-Pv alloy, yielding multiple species on each Mg-Pv sublattice. In the thermodynamic model, we assume the impurity atoms on sub-lattices do not interact with each other. This assumption allows the enthalpy of the whole system to be written as a linear combination of its end-members. Writing the total Gibbs energy of the system in this way allows it to be expressed as a function of $n$ and $y$ at a given pressure and temperature as:

$$G_{TotalOx} = (1-n)H[(Mg_{1-x}Fe_x)SiO_3] + \frac{x+2}{2}nyH[(Mg_{1-x_1}Fe_{x_1})(Si_{1-x_1}\overset{HS}{Fe_{x_1}})O_3] + \frac{x+2}{2}n(1-y)H[(Mg_{1-x_1}Fe_{x_1})(Si_{1-x_1}\overset{LS}{Fe_{x_1}})O_3]$$
$$+\frac{1}{4}(1-n)x\mu(O_2) + (1-n)xH[MgO] - (1-n)xTS_{mag}(Fe^{2+},A) - \frac{x+2}{2}nx_1TS_{mag}(Fe^{3+},A) - \frac{x+2}{2}nyx_1TS_{mag}(Fe^{3+\,HS},B) -$$
$$\frac{x+2}{2}n(1-y)x_1TS_{mag}(Fe^{3+\,LS},B) - (1+\frac{1}{2}nx)TS_{conf}(A) - (1+\frac{1}{2}nx)TS_{conf}(B)$$

(4)

where $G_{TotalOx}$ is the total Gibbs energy for the system with states associated with this oxidation reaction. $S_{conf}(A)$ and $S_{conf}(B)$ are the configurational entropy from the A-site and B-site sublattices of Mg-Pv, respectively. The expressions for $S_{conf}(A)$ and $S_{conf}(B)$ are:

$$S_{conf}(A) = -k[\frac{1-(1-n)x}{1+\frac{1}{2}nx}\ln(\frac{1-(1-n)x}{1+\frac{1}{2}nx}) + \frac{(1-n)x}{1+\frac{1}{2}nx}\ln(\frac{(1-n)x}{1+\frac{1}{2}nx}) + \frac{nx/2}{1+\frac{1}{2}nx}\ln(\frac{nx/2}{1+\frac{1}{2}nx})] \quad (5)$$

$$S_{conf}(B) = -k[\frac{1}{1+\frac{1}{2}nx}\ln(\frac{1}{1+\frac{1}{2}nx}) + \frac{ynx/2}{1+\frac{1}{2}nx}\ln(\frac{ynx/2}{1+\frac{1}{2}nx}) + \frac{(1-y)nx/2}{1+\frac{1}{2}nx}\ln(\frac{(1-y)nx/2}{1+\frac{1}{2}nx})] \quad (6)$$

The $S_{mag}(Fe^{2+},A)$, $S_{mag}(Fe^{3+},A)$, $S_{mag}(Fe^{3+,HS},B)$ and $S_{mag}(Fe^{3+,LS},B)$ are the magnetic entropies of $Fe^{2+}$ in the A site, $Fe^{3+}$ in the A site, high spin $Fe^{3+}$ in the B site and low spin $Fe^{3+}$ in the B site, respectively. These values can be obtained by calculating the degeneracy of the electronic configuration of the $d$-orbital of Fe in different sites and spin states (Sturhahn, 2005; Tsuchiya et al., 2006). The expression for $S_{mag}$ is $k_B\ln[m(2S+1)]$, where $m$ is the electronic configuration degeneracy and $S$ is the iron spin quantum number. The entropy values are given for each as (Sturhahn (2005); Zhang and Oganov (2006)):

| $S_{mag}(Fe^{2+},A)$ | $S_{mag}(Fe^{3+},A)$ | $S_{mag}(Fe^{3+,HS},B)$ | $S_{mag}(Fe^{3+,LS},B)$ |
|---|---|---|---|
| $k_B\ln 10$ | $k_B\ln 6$ | $k_B\ln 6$ | $k_B\ln 6$ |

The equilibrium B-site HS fraction (given by $y$) and $Fe^{3+}$ fraction (given by $n$) can be predicted by minimizing the Gibbs free energy, which in general we will denote with $G$, with respective to $n$ and $y$ by solving the equations:

$$\begin{cases} \frac{\partial G(n,y;T,P)}{\partial n} = 0 \\ \frac{\partial G(n,y;T,P)}{\partial y} = 0 \end{cases} \quad (7)$$

### 2.1.2 Charge disproportionation

Chg. disp. reaction is a spontaneous valence state change of $Fe^{2+}$. The general expression for chg. disp. reaction is:

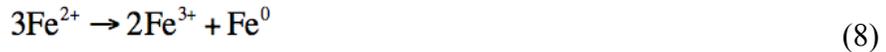
$$3Fe^{2+} \rightarrow 2Fe^{3+} + Fe^0 \quad (8)$$

If there is only $Fe^{2+}$ in the starting Mg-Pv material, $Fe^{3+}$ can be produced by the chg. disp. reaction. The equation for this process in the presence of MgO is:

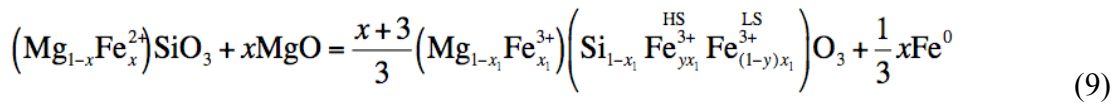
$$(Mg_{1-x}Fe^{2+}_x)SiO_3 + xMgO = \frac{x+3}{3}(Mg_{1-x_1}Fe^{3+}_{x_1})(Si_{1-x_1}\overset{HS}{Fe^{3+}_{yx_1}}\overset{LS}{Fe^{3+}_{(1-y)x_1}})O_3 + \frac{1}{3}xFe^0 \quad (9)$$

where $x_1 = x/(x+3)$. As above for the oxidation reaction, we take $x = 1/8$.

In both oxidation and chg. disp. reaction models, we always use the condition that MgO is excess. This choice is because: (1) in the lower mantle, the molar ratio of MgO/MgSiO$_3$ is about 0.6:1 (Irifune et al., 2010; Kesson et al., 1998; Mao, 1997), so the excess MgO amount is sufficient for the reactions, and (2) in many experiments the starting material is olivine (Mg,Fe)$_2$SiO$_4$, therefore the molar ratio of Fp/Mg-Pv, where Fp is ferropericlass, in these experiments is 1:1 (Auzende et al., 2008; Nakajima et al., 2012; Sinmyo et al., 2008), which again means there is sufficient excess MgO for the reactions.

We follow the same strategy used above for the oxidation reaction to find the Gibbs free energy of the system under the chg. disp. reaction. When a fraction $n$ of the reactants in Eq. (9) are transferred to products, we can write the whole mixed system in the compact form:

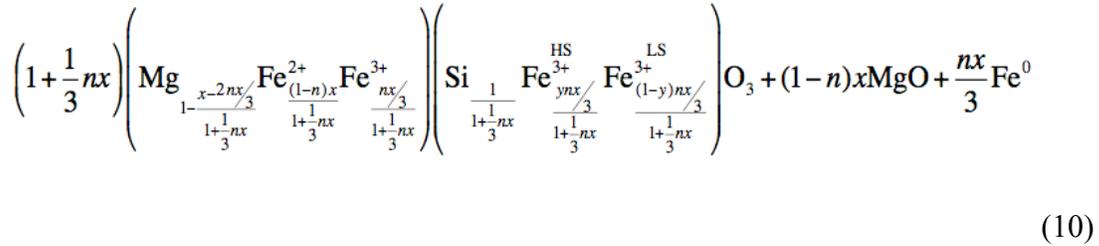

$$\left(1+\frac{1}{3}nx\right)\left(Mg_{\frac{1-x-2nx/3}{1+\frac{1}{3}nx}} Fe^{2+}_{\frac{(1-n)x}{1+\frac{1}{3}nx}} Fe^{3+}_{\frac{nx/3}{1+\frac{1}{3}nx}}\right)\left(Si_{\frac{1}{1+\frac{1}{3}nx}} \overset{HS}{Fe^{3+}_{\frac{ynx/3}{1+\frac{1}{3}nx}}} \overset{LS}{Fe^{3+}_{\frac{(1-y)nx/3}{1+\frac{1}{3}nx}}}\right)O_3 + (1-n)xMgO + \frac{nx}{3}Fe^0$$

(10)

From this expression we can write the total Gibbs energy of the system as:

$$G_{TotalChgDisp} = (1-n)H[(Mg_{1-x}Fe_x)SiO_3] + \frac{x+3}{3}nyH[(Mg_{1-x_1}Fe_{x_1})(Si_{1-x_1}\overset{HS}{Fe_{x_1}})O_3] + \frac{x+3}{3}n(1-y)H[(Mg_{1-x_1}Fe_{x_1})(Si_{1-x_1}\overset{LS}{Fe_{x_1}})O_3]$$
$$+(1-n)xH[MgO] + \frac{xn}{3}H[Fe] - (1-n)xTS_{mag}(Fe^{2+},A) - \frac{x+3}{3}nx_1TS_{mag}(Fe^{3+},A) - \frac{x+3}{3}nyx_1TS_{mag}(Fe^{3+\,HS},B) -$$
$$\frac{x+3}{3}n(1-y)x_1TS_{mag}(Fe^{3+\,LS},B) - \frac{xn}{3}TS[Fe^0] - (1+\frac{1}{3}nx)TS_{conf}(A) - (1+\frac{1}{3}nx)TS_{conf}(B)$$

(11)

where $G_{TotalChgDisp}$ is the total Gibbs energy for the system with states associated with this chg. disp. reaction. Parallel to the oxidation reaction case, we can solve Eq. (7) to find the extent of Fe$^{2+}$ formation (from chg. disp. reaction) $n$ and the B-site HS fraction $y$ as a function of pressure and temperature.

The terms in Eq. (11) are found following the same approach and approximations as described for the oxidation reaction, except that now we must determine the entropy of metallic Fe. For the entropy of metallic Fe we include only magnetic contributions, consistent with the analysis in the oxide phases, and use the theoretical result from the Sommerfeld expansion (Ashcroft and Mermin, 1976),

$$S = \frac{\pi^2 D(E_F)}{3} k_B^2 T \qquad (12)$$

where $D(E_F)$ is the density of states of metallic Fe at Fermi level. We get the $D(E_F)$ as a function of pressure from HSE06 hybrid DFT calculation for non-magnetic hexagonal close packed (hcp) Fe from 20 to 120 GPa.

### 2.1.3 Integrated model

In this section we combine the oxidation reaction model and chg. disp. reaction model together. The approach is the same as with the previous models treated separately. Consider the equilibration process by starting with one mole of $(Mg_{1-x}Fe_x)SiO_3$. When the system comes into equilibrium, $n_1$ mole of ferrous Mg-Pv is transferred to ferric Mg-Pv through the oxidation reaction and $n_2$ mole of ferrous Mg-Pv is transferred to ferric Mg-Pv through the chg. disp. reaction. Then the compact form for the whole system is:

$$\left(1+\frac{1}{2}n_1 x+\frac{1}{3}n_2 x\right)\left(Mg_{\frac{1-(1-n_1-n_2)x}{1+\frac{1}{2}n_1 x+\frac{1}{3}n_2 x}} Fe^{2+}_{\frac{(1-n_1-n_2)x}{1+\frac{1}{2}n_1 x+\frac{1}{3}n_2 x}} Fe^{3+}_{\frac{n_1 x/2 + n_2 x/3}{1+\frac{1}{2}n_1 x+\frac{1}{3}n_2 x}}\right)\left(Si_{\frac{1}{1+\frac{1}{2}n_1 x+\frac{1}{3}n_2 x}} Fe^{3+\,HS}_{\frac{y(n_1 x/2 + n_2 x/3)}{1+\frac{1}{2}n_1 x+\frac{1}{3}n_2 x}} Fe^{3+\,LS}_{\frac{(1-y)(n_1 x/2 + n_2 x/3)}{1+\frac{1}{2}n_1 x+\frac{1}{3}n_2 x}}\right) O_3$$
$$+\frac{1}{4}(1-n_1)xO_2 + (1-n_1-n_2)xMgO + \frac{n_2 x}{3}Fe^0 \qquad (13)$$

The further details of the Gibbs energy expressions are discussed in the SI section 2.

### 2.2 DFT methods

Our ab-initio calculations are performed with the Vienna ab initio simulation package (VASP) based on density functional theory. Projector augmented wave method (PAW) (Blochl, 1994) is used for the effective potential for all the atoms in the system. The PAW potentials we use have $2p^63s^2$ for Mg, $3s^23p^2$ for Si, $2s^22p^4$ for O and $3p^63d^74s^1$ for Fe. A 600 eV energy cutoff is used to make sure the plane wave basis is large enough for converged calculations.

As there are transition metal atoms (Fe) in the system, the normal Local Density Approximation (LDA) and Generalized Gradient Approximation (GGA) functionals often provide inaccurate energetics (Wang et al., 2006). All the calculations in this work are therefore performed with HSE06 hybrid functional (Heyd et al., 2006, 2003; Paier et al., 2006) as implemented in the VASP code. The HSE06 functional has been shown to yield significantly more accurate energetics for transition metal redox reactions than standard LDA or GGA techniques (Chevrier et al., 2010). More quantitatively, Chevrier, et al. (2010) showed that the root mean squared error compared to experiment for a large series of transition metal oxidation reactions is 75 meV per metal atom. Therefore, in the following sections we show the effect of a 75 meV/Fe error on the enthalpies of reaction in our thermodynamic modeling results of $Fe^{3+}$ concentration.

The further details of our DFT method including the endmember supercell setup, equation of state comparison with experiments and the reasons of using HSE06 hybrid functional instead of DFT+U are discussed in the SI section 3.

3. Results
3.1 Integrated model

In general both the oxidation reaction and chg. disp. reaction are possible to occur in either the lower mantle or the experiments. Therefore we first show the integrated model results with both reactions. The model is capable of predicting both $Fe^{3+}/\Sigma Fe$ and $Fe^0$ (metallic Fe) concentrations. The $Fe^0$ concentration reflects the extent of chg. disp. reaction because the $Fe^0$ is produced only from this reaction. Fig. 1 shows the $Fe^{3+}/\Sigma Fe$ ratio and $2[Fe^0]/[Fe^{3+}]$ ratio predicted from the integrated model with respect to $fO_2$ at $P$=100GPa, $T$=2000K. $[Fe^0]$ and $[Fe^{3+}]$ denote the number of moles of $Fe^0$ and $Fe^{3+}$ respectively. $2[Fe^0]/[Fe^{3+}]$ shows the extent of chg. disp. reaction: $2[Fe^0]/[Fe^{3+}]$=1 means only the chg. disp. reaction occurs in the system while $2[Fe^0]/[Fe^{3+}]$ =0 indicates all the $Fe^{3+}$ comes from the oxidation reaction. The relationships between the $fO_2$ of different capsules are from previous experiments (Pownceby and O'Neil, 1994; Frost et al., 2004).

In Fig.1 there is a transition point of $fO_2$ ($fO_2^t$) above which only oxidation reaction occurs and below which only chg. disp. reaction is possible. It is expected that this transition will occur approximately at the $fO_2$ that can oxidize $Fe^0$ to $Fe^{2+}$ (where the $Fe^{2+}$ will occur in ferropericlase), as above that $fO_2$ no $Fe^0$ can exist. We don't use any data from FeO or ferropericlase in constructing our model, so being consistent with $Fe^0$/ferropericlase equilibrium thermodynamics is an important test of the model. In Fig.1 we show the range of the $Fe^0$-ferropericlase equilibrium $fO_2$ derived from experimental data and an ideal solution model (see SI section 5). The grey region in Fig. 1 is the corresponding $fO_2$ range from Fe%=5 to Fe%=40 in ferropericlase calculated from ideal solution model. We can see that the transition point occurs at the $fO_2$ where $Fe^0$ is in equilibrium with $(Mg_xFe_{1-x})O$ (x≈0.36). This result implies that under the $(Mg_{0.875},Fe_{0.125})SiO_3$ compositional condition for Mg-Pv, if we include Fp in the system, the Fe% in Fp is about 36 when we have $Fe^0$ formation and the corresponding equilibrium $fO_2$ is the $fO_2^t$ we show in Fig.1. For $fO_2$ below $fO_2^t$ the equilibrium will be for lower Fe% in Fp, and an $fO_2$ of about 11.6 corresponds to about Fe%=20 in the Fp, consistent with values often suggested for the lower mantle (Hirose, 2002). From Fig. 1 we see that our model predicts a range of $fO_2$ above where Fp with Fe% consistent with the lower mantle is stable against forming $O_2$ gas and $Fe^0$ and below where $O_2$ gas starts to be consumed by oxidation of $Fe^{2+}$ to $Fe^{3+}$ in Mg-Pv. This range is therefore where our model (which contains no Al) would predict the $fO_2$ of the lower mantle to reside, somewhere around 11.6. The ability to define this range consistently for both Mg-Pv and Fp, despite the model being developed without any explicit *ab initio* calculations on the Fp system, supports the accuracy of our thermodynamic model.

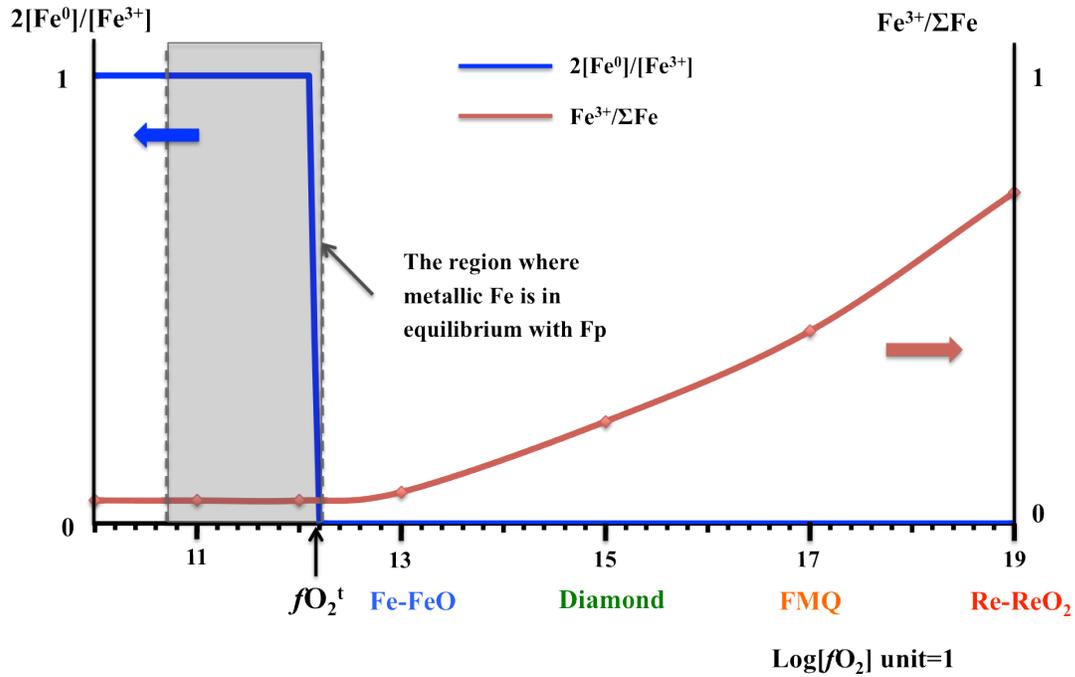

**Fig. 1** $2[Fe^0]/[Fe^{3+}]$ and $Fe^{3+}/\Sigma Fe$ with respect to $fO_2$ at $T = 2000K$, $P = 100$ GPa. The blue curve corresponds to $2[Fe^0]/[Fe^{3+}]$ ratio and the red curve corresponds to $Fe^{3+}/\Sigma Fe$. The $fO_2$ of different capsules are from experiments (references in text). The grey region is the $Fe^0$-ferropericlase equilibrium $fO_2$ at Fe% = 5 (left boundary) and Fe% = 40 (right boundary).

In the following sections we apply our thermodynamic model for two different situations. The first situation is to model laboratory experiments. In laboratory experiments the $fO_2$ is usually buffered by the capsules and has values higher than $fO_2^t$. As can be seen from Fig. 1, there is no metallic Fe in these higher $fO_2$ and therefore no chg. disp. reaction. Therefore, we will use our oxidation model to study $Fe^{3+}$ content at different $P$, $T$ and capsules conditions (see section 3.2). The second situation is to model conditions in the lower mantle. For the lower mantle we assume the stoichiometry consists of a nominally ferrous Mg-Pv and ferropericlase system (i.e., $(Mg_{1-x}Fe_x)SiO_3+(Mg_{1-y}Fe_y)O$ and all Fe is $Fe^{2+}$) from pyrolitic and hartzburgitic compositions, and that this stoichiometry is fixed. We assumed that there is no external free oxygen gas or negligible amount of other oxidizing agents to oxidize $Fe^{2+}$ in the lower mantle. Therefore, in our study, the only possible mechanism to produce $Fe^{3+}$ is the chg. disp. reaction and this reaction will be the focus of the modeling (see section 3.3). Under these assumptions appropriate for modeling the lower mantle the $fO_2$ will be below $fO_2^t$ and $Fe^0$ can be stabilized.

**3.2 Oxidation reaction**

We calculate the concentration of $Fe^{3+}$ produced by the oxidation reaction from 20 to 120 GPa at 2000, 3000, and 4000 K with $fO_2$ set by the Re-ReO$_2$ capsule (higher $fO_2$) and diamond capsule (lower $fO_2$) buffers. All the details of capsules and their $fO_2$ values as a function of pressure are given in the SI section 1. The results of $Fe^{3+}$ concentration in different capsules are shown in **Fig. 2.** In the Al-free system, the oxidation reaction produces a significant amount of $Fe^{3+}$. The general tendency of $Fe^{3+}/\Sigma Fe$ is that it first decreases up to 40 GPa and then increases. At 2000 K, when the pressure is relatively low, between 20 and 60 GPa, $Fe^{3+}/\Sigma Fe$ is ~0.5 in the Re-ReO$_2$ capsule (higher $fO_2$) and is about 0.08 in diamond capsule (lower $fO_2$). When the pressure increases to 100 GPa, $Fe^{3+}/\Sigma Fe$ is ~0.7 in Re-ReO$_2$ and ~0.2 in diamond capsule. The ratio is also quite sensitive to temperature and the $Fe^{3+}$ concentration increases when the temperature increases. We also use the model to study the spin transition of $Fe^{3+}$ in the B site of Mg-Pv (**Fig. 3**). Our thermodynamic model indicates a gradual spin transition for the $Fe^{3+}$ in the B site of Mg-Pv. At 2000 K, the spin transition occurs from 30 to 50 GPa. Previous experimental results showed that the spin transition range is 40~60 GPa (Catalli et al., 2010) and simulation results predicted that it is 40~70 GPa (Hsu et al., 2011), which are consistent with this work.

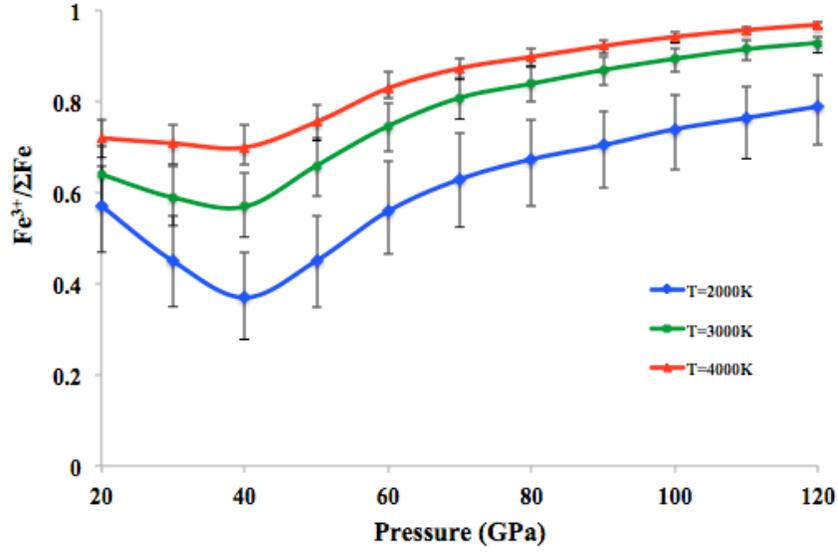

(a)

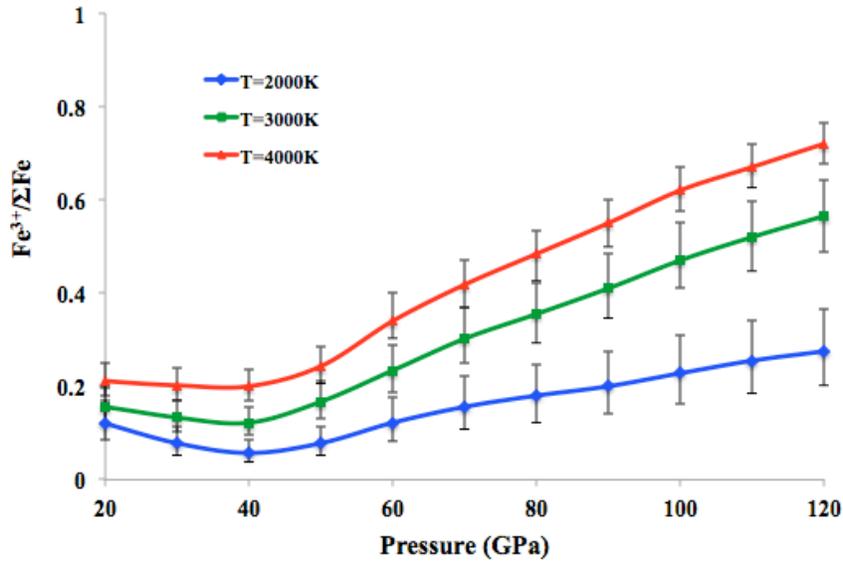

(b)

**Fig. 2** $Fe^{3+}/\Sigma Fe$ due to the oxidation reaction (**Eq. (1)**) at high pressures and temperatures. (a) $Fe^{3+}/\Sigma Fe$ in the Re-ReO$_2$ capsule (higher $fO_2$) and (b) $Fe^{3+}/\Sigma Fe$ in the diamond capsule (lower $fO_2$). The error bars incorporate the variance of HSE06 hybrid functional result on transition metal oxides, which is taken as ±75 meV/Fe (Chevrier et al., 2010).

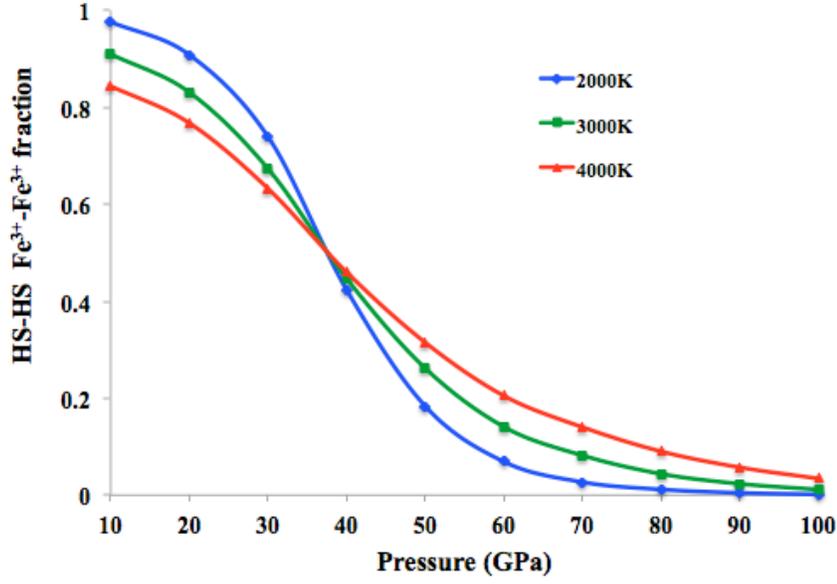

**Fig. 3** Spin transition of $Fe^{3+}$ in the B site of ferric Mg-Pv. We show the spin transition with respect to pressures at different temperatures. We denote the fully high-spin state as HS-HS, where the first HS represents the spin state of the $Fe^{3+}$ in the A site and the second HS represents the spin state of the $Fe^{3+}$ in B site in ferric Mg-Pv. The state transforms to HS-LS as the systems undergoes the spin transition.

To better understand what controls the $Fe^{3+}$ concentration, we analyze the energetics of Eq. (4) by separating the Gibbs energy change of the reaction (Eq. (1)) into enthalpy change and entropy change. In the entropy we do not include the configurational entropy as the $S_{config}$ is concentration dependent (it will be discussed later in the next paragraph). **Fig. 4** shows the contribution of each term in $\Delta G$ of Eq. (1). In **Fig. 4** (a), $\Delta G$ first increases up to 40 GPa and then decreases at higher pressures. The trend in $\Delta G$ creates the convex shape of $Fe^{3+}/\Sigma Fe$ versus pressure in **Fig. 2**. **Fig. 4** (a) also shows that the concave shape of $\Delta G$ is related to the change in slope of $\Delta H$ at ~40 GPa. In Fig. 4(b), $\Delta H$ is further split into $P\Delta V$ and $\Delta U$ ($\Delta H = \Delta U + P\Delta V$). It is clear that the $P\Delta V$ also has a slope change at ~40 GPa and that the concave shape of $Fe^{3+}/\Sigma Fe$ is due to this behavior of the $P\Delta V$ term. This slope change in the $P\Delta V$ term is caused by the spin transition of B site $Fe^{3+}$ from high to low-spin state. Below 40 GPa, the majority spin configuration is the HS-HS ferric Mg-Pv, and $P\Delta V$ increases with pressure going up. We denote the fully high-spin state as HS-HS, where the first HS represents the $Fe^{3+}$ spin state in the A-site and the second represents $Fe^{3+}$ spin state in B-site in ferric Mg-Pv. When the pressure is ~40 GPa, HS-HS ferric Mg-Pv begins to transform into HS-LS Mg-Pv leading to a volume collapse, so the $P\Delta V$ increase rate is reduced significantly around the spin transition pressure. Then, as the pressure increases, all the ferric Mg-Pv is in HS-LS state and the curve is again smooth. So the change in slope is generated in the $P\Delta V$ and also in the $\Delta H$ curve. Based on the above analysis, we propose that it is the volume reduction of the spin transition of B site $Fe^{3+}$ that causes the convex shape of the $Fe^{3+}/\Sigma Fe$.

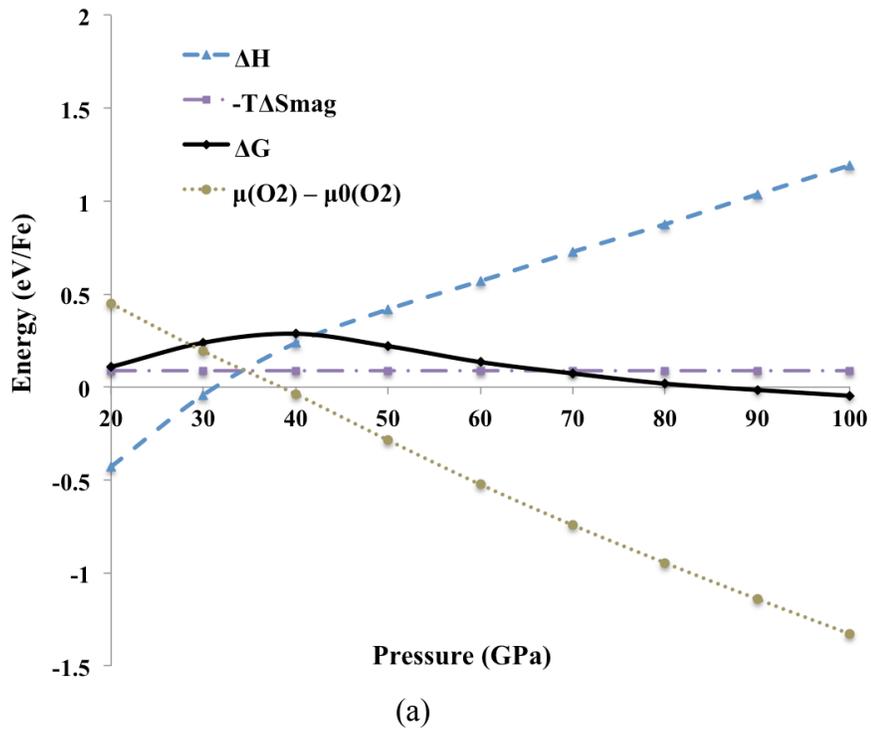

(a)

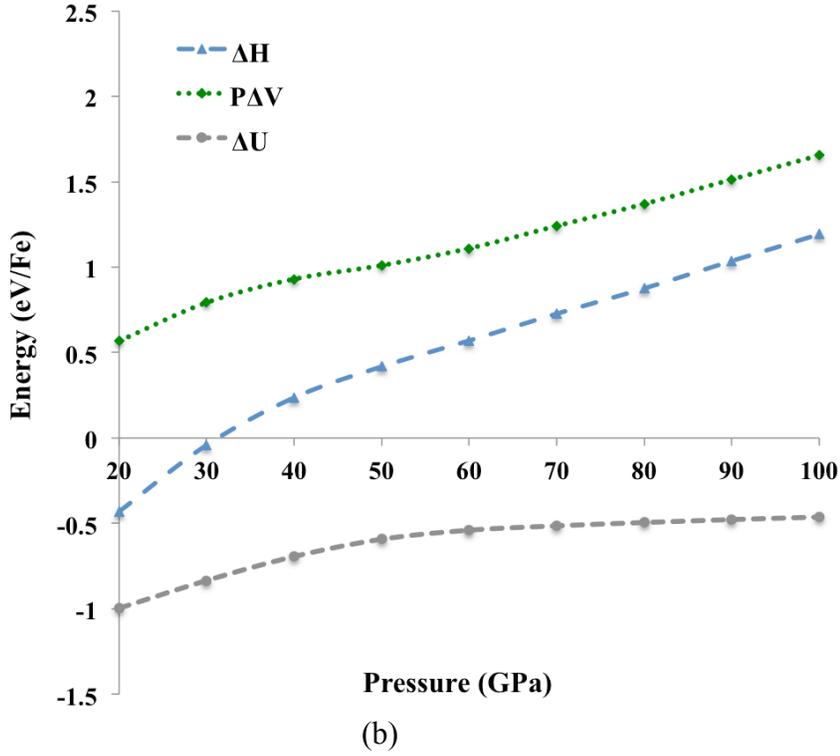

(b)

**Fig. 4** Changes in thermodynamic quantities with respect to pressure between the right hand side and left hand side of the oxidation reaction in Eq. (1). The lines show change in Gibbs free energy ($\Delta G$), enthalpy ($\Delta H$), internal energy ($\Delta U$), temperature times magnetic entropy ($-T\Delta S_{mag}$), pressure times volume ($P\Delta V$), and the chemical potential difference between $O_2$ at ($T,P$) ($\mu(O_2)$) with respect to its reference value at $T = 300$ K, $P = 1$ atm ($\mu_0(O_2)$). Terms are related by $\Delta G = \Delta H - T\Delta S_{mag} = \Delta U + P\Delta V - T\Delta S_{mag}$. All values are shown at $T = 2000$ K for the Re-ReO$_2$ capsule. The energy is labeled by eV/Fe, which means the system unit we choose has one Fe atom in it. (a) $\Delta G$ and its contributions. (b) $\Delta H$ and its contributions.

To understand what sets the specific concentration of $Fe^{3+}$ we show in **Fig. 5** the changes in value of each of the terms contributing to $G_{TotalOx}$ (including configurational entropy), as well as the total change in $G_{TotalOx}$ value, as the Fe oxidation occurs. These values are determined from Eq. (4) and take $n = 0$ (pure $Fe^{2+}$) as the reference (all values are zero for this case). The data are shown as a function of $n$ and for the spin state ($y$) which minimizes the total $G_{TotalOx}$ at fixed $n$, all for 2000 K, 40 GPa, and $fO_2$ given by Re-ReO$_2$. We note that $n$ gives the amount of $Fe^{3+}$, or equivalently, represents the amount of the reaction in Eq. (1) that has taken place at equilibrium (amount of ferrous Mg-Pv oxidized and transformed to ferric Mg-Pv component). It can be seen clearly that as $n$ increases the enthalpy actually increases, demonstrating that formation of ferric Mg-Pv through the reaction in Eq. (1) is enthalpically unfavorable under these conditions. However, enthalpy increase competes with a decrease in $G_{TotalOx}$ from the configurational entropy contribution, which is the term is $G_{TotalOx}$ primarily driving the formation of $Fe^{3+}$. The dominant role of configurational entropy explains why $Fe^{3+}/\Sigma Fe$ in Mg-Pv is predicted to be sensitive to temperature.

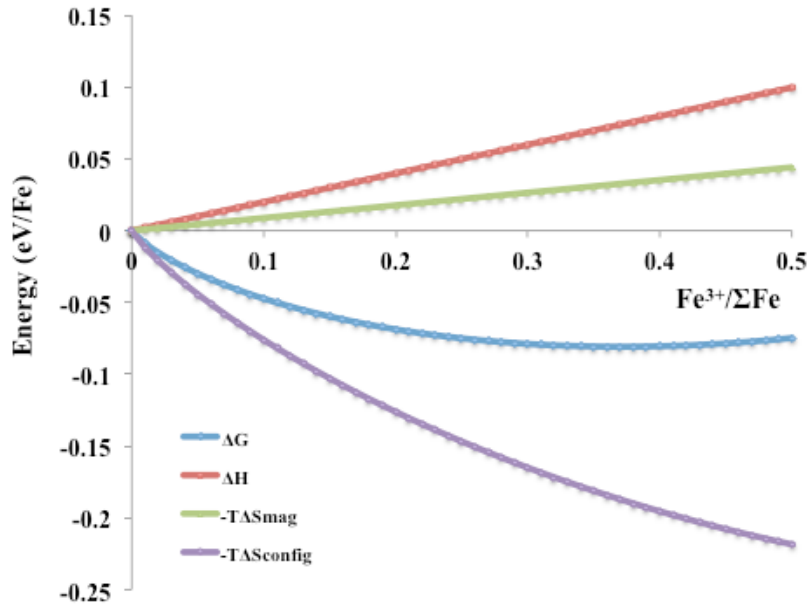

**Fig. 5** Changes in total and contributions to the Gibbs free energy as a function of amount of ferrous Fe oxidized to ferric iron ($Fe^{3+}/\Sigma Fe$) under the oxidation reaction. Note that the variable $n$ from Eq. (4) is equal to the $Fe^{3+}/\Sigma Fe$ ratio in this figure. $\Delta G(Fe^{3+}/\Sigma Fe) = \Delta H(Fe^{3+}/\Sigma Fe) - T\Delta S_{mag}(Fe^{3+}/\Sigma Fe) - T\Delta S_{config}(Fe^{3+}/\Sigma Fe)$. Terms are taken from Eq. (4) and evaluated at $T = 2000$ K, $P = 40$ GPa, and $fO_2$ given by Re-ReO$_2$. The results show that the increase in configurational entropy drives the Fe oxidation.

**3.3 Charge disproportionation**

In this section we calculate the $Fe^{3+}$ produced by the chg. disp. reaction in Eq. (9) from 20 to 120 GPa at 2000, 3000, and 4000 K. This mechanism was proposed by some previous studies to be a major source of valence state change of Fe under the lower mantle conditions (Frost et al., 2004; Lauterbach et al., 2000; Zhang and Oganov, 2006), where the $fO_2$ is expected to be very low. The chg. disp. reaction also produces metallic Fe in the lower mantle (Frost et al., 2004). Previous simulation (Zhang and Oganov, 2006) predicted chg. disp. reaction in both Al-free and Al-bearing system and experimental (Frost et al., 2004; Lauterbach et al., 2000; Grocholski et al., 2009) studies suggested Fe disproportionation in Al-bearing systems. But there were some significant discrepancies between experiments and simulation. In particular, previous calculations of the enthalpy change of the chg. disp. reaction (Eq. (9)) for both Al-bearing and Al-free cases predicted a negative value of -1.13 eV/Fe and -1.03 eV/Fe respectively, suggesting that the chg. disp. reaction goes to near completion and almost all $Fe^{2+}$ will be transformed to $Fe^{3+}$ (i.e., $Fe^{3+}/\Sigma Fe \approx 1$). However, the values of $Fe^{3+}/\Sigma Fe$ in experiments are at most about 0.2~0.3 for Al-free Mg-Pv and 0.6 for Al-bearing Mg-Pv. The discrepancy suggests that there is significant uncertainty about the driving forces and extent of the chg. disp. reaction in Mg-Pv.|

In this work, the predicted $Fe^{3+}$ content by chg. disp. reaction is shown in **Fig. 6**. We emphasize that these predictions are independent of $fO_2$ provided $fO_2$ is less than $fO_2^t$ in Fig. 1. The extent of chg. disp. reaction is much lower than that predicted from the oxidation reaction. From 20 to 120 GPa, the $Fe^{3+}$ fraction is only around 0.01~0.07 at 2000 K. Even at 4000 K, which is likely higher than the temperature for most of the lower mantle, the $Fe^{3+}$ concentration from chg. disp. reaction is still quite limited, predicted to be about 0.08~0.15. Similarly to the oxidation reaction, we predict that the enthalpy of the chg. disp. reaction is positive and that the $Fe^{3+}$ formation is driven by gains in configurational entropy, which implies that $Fe^{3+}/\Sigma Fe$ is sensitive to temperature.

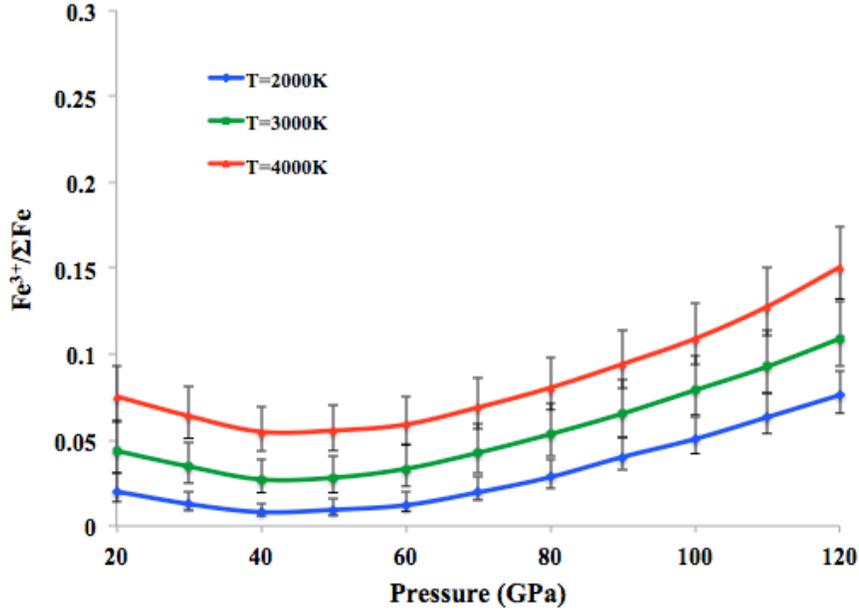

**Fig. 6** $Fe^{3+}/\Sigma Fe$ due to the chg. disp. reaction only at different pressures and temperatures. The error bar incorporates the variance of HSE06 hybrid functional result on transition metal oxides, which is taken as ±75 meV/Fe (Chevrier et al., 2010).

Similar to the results in the oxidation reaction, the $Fe^{3+}$ concentration first decreases and then increases with pressure. As with the oxidation reaction, the turning point is at ~40 GPa and can be explained by the B-site spin transition. **Fig. 7** shows the energetics of Eq. (9) by splitting the Gibbs energy change into $\Delta H$ (=$\Delta U$+ $P\Delta V$) and $-T\Delta S_{mag}$. The result is similar to **Fig. 4** in that a change in slope in the $P\Delta V$ curve leads to a convex shape of $\Delta H$ and $\Delta G$. Before the spin transition, the reaction in of Eq. (9) causes a volume increase, after the spin transition it leads to a volume decrease, where the change is due to smaller size of LS with respect to HS $Fe^{3+}$ in the final state. As $\Delta V$ changes from positive to negative the value of $P\Delta V$ changes sign as well. The shift in sign of this term changes $\Delta H$ from increasing to decreasing with pressure, which then leads to the convex shape of $Fe^{3+}/\Sigma Fe$ as a function of pressure.

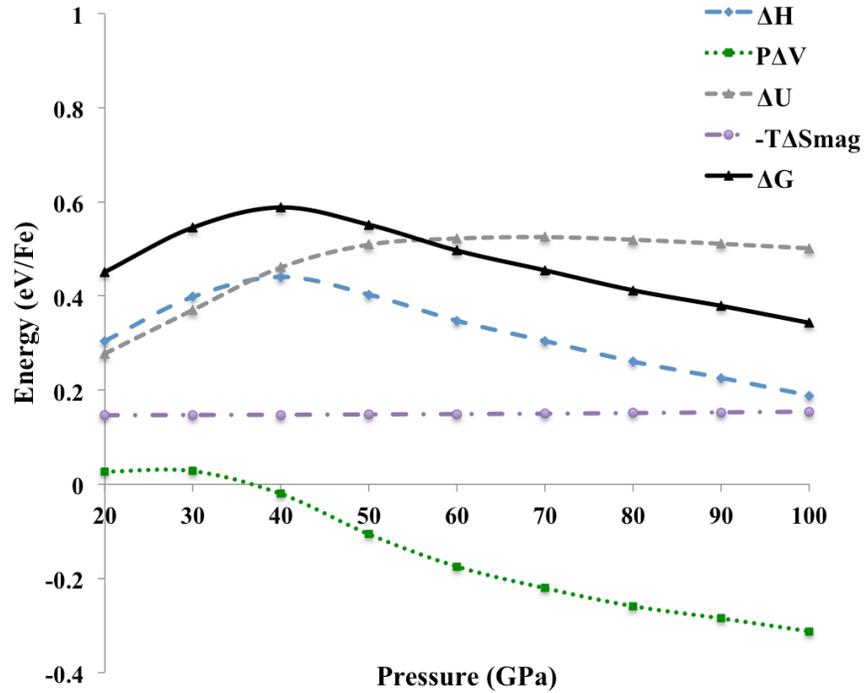

**Fig. 7** Changes in thermodynamic quantities with respect to pressure between the right- and left-hand sides of the chg. disp. reaction in Eq. (9). The lines show change in Gibbs free energy ($\Delta G$), enthalpy ($\Delta H$), internal energy ($\Delta U$), temperature times magnetic entropy ($-T\Delta S_{mag}$), and pressure times volume ($P\Delta V$). Terms are related by $\Delta G = \Delta H - T\Delta S_{mag} = \Delta U + P\Delta V - T\Delta S_{mag}$. All values are shown at $T = 2000$ K. The energy is labeled by eV/Fe, which means the system unit we choose has one Fe atom in it.

We also calculate the Al-free $Fe^{3+}/\Sigma Fe$ contribution from chg. disp. reaction under specifically lower mantle conditions, as shown in in **Fig. 8.** We set the lower mantel $T$-$P$ by using the geotherm by Brown and Shankland (1981).

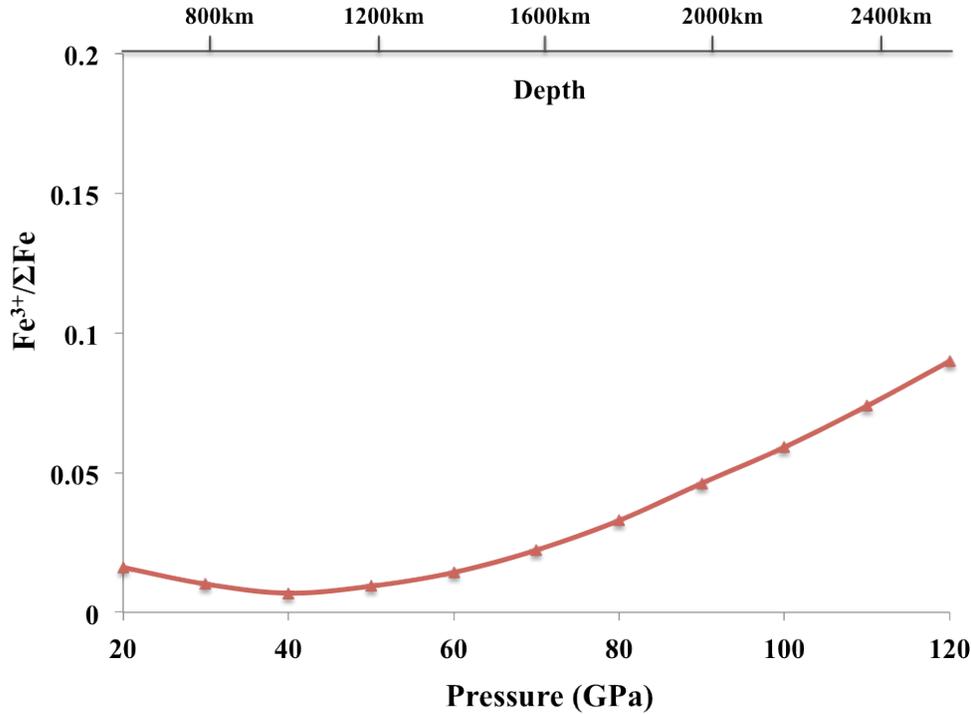

**Fig. 8** Calculated $Fe^{3+}/\Sigma Fe$ in Mg-Pv with respect to pressures from chg. disp. reaction under lower mantle geotherm (Brown and Shankland (1981)) *P-T* condition.

## 4. Discussion

### 4.1 Validation of the model

Experimental studies have shown that $Fe^{3+}/\Sigma Fe$ is about 0.2~0.3 in the Re-ReO$_2$ capsule and close to zero in the diamond capsule in Al-free Mg-Pv at 25~30 GPa and 2000 K (Lautherbach et al., 2000; Frost et al., 2002). In our modeling result, $Fe^{3+}/\Sigma Fe$ is about 0.4-0.5 in the Re-ReO$_2$ capsule and 0.08-0.1 in the diamond capsule at the same *P-T* range.

Our modeling results are slightly higher than the experimental results for $Fe^{3+}/\Sigma Fe$. One source of this discrepancy may be kinetic limitations in the experiments. Whereas our modeling work is based on the thermodynamic equilibrium (minimization of the total Gibbs energy), in experiments there may be kinetic effects limiting the formation of $Fe^{3+}$. We predict that the dominant mechanism for forming $Fe^{3+}$ in many experiments is the oxidation reaction of $Fe^{2+}$ in the A-site of Mg-Pv to $Fe^{3+}$ in both A- and B-sites in Al-free Mg-Pv. This reaction requires diffusion of some Fe ions from A to B sites. The kinetic barrier for the $Fe^{3+}$ diffusion between different sites and the associated Mg vacancy migration could prevent some $Fe^{2+}$ being oxidized to $Fe^{3+}$, which would decrease $Fe^{3+}/\Sigma Fe$. According to the two reactions we consider in this work Eq. (1) and Eq. (9), when $Fe^{3+}$ enters the Mg-Pv by charge coupled substitution, diffusion of a Mg vacancy must also occur. According to Ammann et al. (2010), the kinetic barriers for Mg vacancy mediated migration at 30 GPa is about 4.0 eV. The diffusion coefficient can be approximately estimated from this barrier using the relationship: $D = a^2 \upsilon \times \exp(-\frac{E_{barrier}}{kT})$, where $a$ is the hop length and $\upsilon$ is the attempt frequency. Taking reasonable values of $a$ = 3Å which is about the Mg-Mg distance in Mg-Pv and $\upsilon = 5 \times 10^{12}$ Hz we can estimate $D$ and thereby assess kinetic effects. We assume that the associated transport of Mg vacancy requires diffusion of relevant species at least to the grain boundary between MgO and Mg-Pv. This assumption is supported by the observation that when we have a Mg vacancy created in the system, the vacancy has to diffuse to the MgO/Mg-Pv boundary to react with MgO and compensate the vacancy, as shown in **Eq. (1)** and **Eq. (9)**. From the TEM images of Mg-Pv coexisting with MgO (Frost et al., 2004; Irifune et al., 2010), the length scale of the Mg-Pv grain is about 5 μm, so the total diffusion length for Mg vacancies migrating to the grain boundary is about $L$ = 2.5 μm. The diffusion time is approximately given by $t = L^2/6D$. Therefore at 30 GPa and 2000 K, the total diffusion time for Mg vacancies to the grain boundary is about 8 hrs. However, in some relevant experiments to date the Mg-Pv sample was under high $P$-$T$ for less than 8 hrs, for example: 2 hrs (Sinmyo et al., 2008), 2 hrs (Lauterbach et al., 2000). Although the estimates for diffusion times and associated time to produce $Fe^{3+}$ are very approximate, this analysis suggests that the experimental duration may not be sufficiently long to allow the system to evolve into complete thermal equilibrium. Furthermore, the rate of the oxidation reaction will slow down as more $Fe^{3+}$ forms and the configurational entropy driving force becomes weaker. From Fig. 5 it is clear that the gain in Gibbs energy from forming additional $Fe^{3+}$ becomes very minor as the concentration approaches its equilibrium value.

Another likely source for discrepancies between the model and the experiment is uncertainties in the ab initio energetics and/or equations of state. To obtain values closer to experiment, e.g., $Fe^{3+}/\Sigma Fe = 0.25$ in the Re-ReO$_2$ capsule and 0.05 in the diamond capsule at 2000 K and 30 GPa, our reaction energies would have to be increased by about 0.17 eV/Fe and 0.10 eV/Fe, which correspond to a change in reaction volume of 0.89 Å$^3$/Fe and 0.52 Å$^3$/Fe, respectively. While these energy errors are larger than the 75 meV/Fe we used in our estimated error bars, they are not outside the range of errors sometimes seen in transition metal oxide calculations (Chevrier et al., 2010).

Overall, it is possible that both errors in model energetics and some kinetic limitations in the experiments lead to the observed discrepancies. However, the level of agreement for $Fe^{3+}/\Sigma Fe$, especially in light of the excellent agreement between the calculated and measured equation of state parameters (see Table S4), suggests that the model is robust enough to make useful prediction for mechanisms, trends, and perhaps even quantitative values for the $Fe^{3+}$ formation.

Zhang and Oganov (2006) performed a pioneering series of ab-initio reaction enthalpy calculations for the chg. disp. reaction given in Eq. (9) and found a large negative $\Delta H$, suggesting nearly complete chg. disp. However, our thermodynamic modeling indicates that the configurational entropy drives the reactions and the $\Delta H$ is positive (Figs 4, 5, 7). This difference comes from the fact that Zhang and Oganov (2006) used standard GGA to calculate the enthalpy difference of the chg. disp. reaction. Their paper was the first DFT simulation work to provide us with insights about the formation energetics of $Fe^{3+}$ from $Fe^{2+}$, but they faced many challenges in accurately treating transition metal oxide energetics. In particular, techniques for treating correlated electron systems with the HSE06 hybrid functional hadn't been well studied and the various $U$ values for Fe in Mg-Pv were not yet known then. Based on the previous discussion about the accuracy of HSE06 functional and the work of Chevrier et al., (2010), we believe that our HSE06 hybrid functional calculations are significantly more reliable that the previous standard GGA studies of Zhang and Oganov (2006).

Diamond-anvil cell studies have suggested similar level of $Fe^{3+}/\Sigma Fe$ in Mg-Pv as that reported in multi-anvil press studies at greater depth conditions. Whereas multi-anvil press experiments controlled $fO_2$ of the samples through various methods (Frost et al., 2004; Lautherbach et al., 2000), the exact $fO_2$ was unknown in almost all the diamond-anvil cell experiments, except for Grocholski et al (2009, GRL) where metallic iron powder was added in an attempt to regulate $fO_2$ to reducing conditions. Our study uses the $fO_2$ of the capsule materials to approximate the $fO_2$ in the sample chamber and suggests that $Fe^{3+}/\Sigma Fe$ can be highly sensitive to $fO_2$ at least in Al-free Mg-Pv and therefore it is important to make attempts to control and measure $fO_2$ in diamond-anvil cell synthesis of Mg-Pv.

**4.2 Geophysical implications of the modeling**

Our thermodynamic model in this work only focuses on the Al-free Mg-Pv and therefore can be used to model lower-mantle regions with low Al content. In pyrolite, which is believed to be average composition of the mantle, at 20~25 GPa, Irifune et al. (2010) showed that most Al remains in majoritic garnet and Mg-Pv contains little Al. Therefore, our model predicts that in the uppermost lower mantle, where Al content is low in Mg-Pv, the formation of $Fe^{3+}$ is unfavorable and therefore the $Fe^{3+}$ concentration in Mg-Pv and the content of metallic Fe are expected to be very low. We predict that the upper bound of equilibrium $Fe^{3+}/\Sigma Fe$ in Mg-Pv is about 0.015 at 25 GPa at mantle temperature, as indicated from **Fig. 8.**

Harzburgite, a depleted mantle composition, is expected to exist in the mantle, including the lower mantle, through subduction of oceanic lithosphere. It contains much smaller amount of Al than pyrolite and therefore Mg-Pv in this composition may contain substantially lower Al (Xu et al., 2008). Our results suggest such heterogeneities will contain Mg-Pv with very little $Fe^{3+}$ and metallic iron.

One of the most intriguing observations for geophysics in our study is that $Fe^{3+}/\Sigma Fe$ in Mg-Pv may not stay constant with depth in the lower mantle. Within our model this is not due to changes in oxygen content but changes in the extent of the chg. disp. reaction.

In some recent geochemical models, $Fe^{3+}/\Sigma Fe$ in Mg-Pv has been a central property in explaining metal-silicate partitioning during core formation and elevation of upper-mantle $fO_2$ to current level (Wood et al., 2006; Frost and McCammon, 2008; McCammon, 2005). However, in these models, $Fe^{3+}/\Sigma Fe$ in Mg-Pv is assumed to be constant at different depths in the lower mantle. Our study reveals that effects from spin transition and temperature can result in complex radial variations in $Fe^{3+}/\Sigma Fe$ (at least in Al-free Mg-Pv) due to changes in the extent of the chg. disp. reaction, even if the oxygen content of the lower mantle remains constant. Although the magnitude is small, our study suggests that the mid mantle may contain less $Fe^{3+}$ in Mg-Pv than the rest of the mantle. The lowermost mantle is a thermal boundary layer and temperature is expected to increase rapidly with depth from ~2500 K to ~4500 K within 200-400 km depth interval. Whereas Mg-Pv may transform to the post-perovskite phase in the lowermost mantle, warm regions in the lowermost mantle will be still dominated by Mg-Pv due to the strong positive Clapeyron slope of the post-perovskite boundary (Shim, 2008; Tateno, 2009). Large Low Shear Velocity Provinces (LLSVPs) have been identified in the lowermost mantle (Garnero and McNamara, 2008). The low velocities of the regions are often attributed to enrichment in Fe and elevated temperature in the heterogeneity (McNamara, 2005). Our models predict that $Fe^{3+}$ concentration in Mg-Pv may reach the maximum due to the possible thermal effects in LLSVP. The elevation of the $Fe^{3+}$ concentration may alter the physical properties of Mg-Pv, possibly affecting seismic properties (such as density and bulk modulus) and transport properties (such as thermal and electrical conductivities) (Catalli et al., 2011; Goncharov et al., 2008), and therefore may have impact on the dynamic stability of LLSVPs in the lowermost mantle.

## 5. Conclusion

We use an ab-initio based thermodynamic model to study the origin and extent of $Fe^{3+}$ formation in Al-free Mg-Pv. The model uses the accurate HSE06 functional to treat the correlation effects associated with the Fe and finds good agreement with the equation of state and the spin transition pressure of $Fe^{3+}$ in the B site of ferric Mg-Pv reported in previous models and experiments. We consider two mechanisms for the $Fe^{3+}$ formation: (a) the oxidation reaction and (b) the charge disproportionation reaction. For both reactions this work suggests that, contrary to previous models, the configurational entropy is the driving force to form $Fe^{3+}$ and that $Fe^{3+}$ is not favored by enthalpy gains under lower-mantle or typical experimental conditions. This dominance of entropy results in a strong temperature dependence of $Fe^{3+}/\Sigma Fe$. The oxidation reaction is the dominant mechanism producing $Fe^{3+}$ in laboratory experiments with buffering from capsules that control the $fO_2$ at a relative high level compared to the Fe – ferropericlase equilibrium. Simulated $Fe^{3+}/\Sigma Fe$ in the Re-ReO$_2$ and diamond capsules is qualitatively consistent with the experiments and discrepancies may be in part due to only partial equilibration during the experiments. The model predicts that the charge disproportionation reaction will yield an $Fe^{3+}/\Sigma Fe$ ratio of about 0.01~0.07 at lower-mantle conditions in Al-free Mg-Pv. This result suggests $Fe^{3+}$ concentration is very limited in the absence of dissolved Al in the Mg-Pv. Thus we expect limited $Fe^{3+}$ for Mg-Pv in the uppermost region of the lower mantle (25~30 GPa) and Mg-Pv in harzburgite throughout the lower mantle, as both contain very little Al. Our results also suggest that $Fe^{3+}/\Sigma Fe$ in Al-free Mg-Pv does not remain constant in the lower mantle but may be low in the mid mantle and reach a maximum at the core-mantle boundary region.

**Acknowledgement**

Morgan and Shim gratefully acknowledge support from the National Science Foundation (NSF) for this work (NSF-EAR1316022). Computing resources in this work benefitted from the use of the Extreme Science and Engineering Discovery Environment (XSEDE), which is supported by National Science Foundation grant number OCI-1053575.

**Supplemental Information:**

**1. $fO_2(T,P)$ of different capsules**

The modeling in this work requires knowing the $fO_2$ in different capsules (Re-ReO$_2$, diamond) and under lower mantle conditions at different temperatures and pressures. From the work of (Campbell et al., 2007), we can obtain the Re-ReO$_2$ capsule isothermal $fO_2$ at $P$=0GPa, 20GPa, 40GPa for $T$=1500K and 2000K. However, we have to develop a method to extrapolate the $fO_2$ up to 120GPa and to T=3000K and 4000K.

Following (Campbell et al., 2009), for an oxidation reaction of the form

$$M + \frac{x}{2}O_2 = MO_x \tag{S1}$$

the effect of pressure on $fO_2$ depends on the volume difference ($\Delta V$) between oxide and metal through

$$\left.\frac{\partial \ln[f(O_2)]}{\partial P}\right|_T = \frac{2}{xRT}\Delta V \tag{S2}$$

Eq. (S2) here correspond to the Eq. (3) in (Campbell et al., 2009).

In another section of (Campbell et al., 2007), the $\Delta V(P)$ curve of the [ReO$_2$-Re] pair is shown and it is clear that $\Delta V(P)$ is well-approximated by a linear function of $P$ at a given temperature. We therefore assume we can write

$$\Delta V(P) = A \times P + B \tag{S3}$$

where $A$ and $B$ are undetermined constants. Integrating Eq. (S2) with the form in Eq. (S3) yields

$$\ln[f(O_2)]_T = A_1 \times P^2 + B_1 \times P + C_1 \tag{S4}$$

which has three undetermined parameters, $A_1$, $B_1$, and $C_1$. We can now use the values for $fO_2$ at $P$=0, 20 and 40GPa on the $T$=2000K isothermal line to fit the $A_1$, $B_1$ and $C_1$. We then us this fitted expression to extrapolate the $fO_2$ at higher pressures.

For the temperature dependence we take a similar approach, noting that there is an approximately linear relationship between $\ln[f(O_2))]$ and $1/T$ at a fixed pressure for a metal-metal oxide pair according to Fig 6. (a) and (b) in (Campbell et al., 2009). Therefore, at a certain pressure, we can approximate:

$$\ln[f(O_2)]_T = A_1 \times P^2 + B_1 \times P + C_1 \tag{S5}$$

where $A_2$ and $B_2$ are undetermined constant. As we already have the $fO_2$ of Re-ReO$_2$ capsule at $T$=1500K and $T$=2000K, it is straightforward to extrapolate to higher temperatures.

We can now use these expressions as a guide to find the general expression of Log[$f$(O$_2$)] as a function of $T$ and $P$. As the Log[$f$(O$_2$)]|$_P$ is a linear function of 1/$T$, for convenience we write Log[$f$(O$_2$)]($P$, 1/$T$) and consider a Taylor expansion around $P$=0 and 1/$T$=0 (to third order) as:

$$Log[f(O_2)](P,\frac{1}{T}) = L_0 + L_1 \times P + L_2 \times \frac{1}{T} + L_3 \times P^2 + L_4 \times \frac{P}{T} + L_5 \times \frac{1}{T^2} + L_6 \times P^3 + L_7 \times \frac{P^2}{T} + L_8 \times \frac{P}{T^2} + L_9 \times \frac{1}{T^3}$$

$$\tag{S6}$$

Based on equation S4 and S5, we can accurately represent $P$ and 1/$T$ dependence without terms of the form $P^3$, $(1/T)^2$ and $(1/T)^3$. Therefore, we can take $L_5$, $L_6$, $L_8$ and $L_9$ in S6 as equal to zero. Rewriting Eq. (S6) with this further approximation yields:

$$Log[f(O_2)](P,\frac{1}{T}) = L_0 + L_1 \times P + L_2 \times \frac{1}{T} + L_3 \times P^2 + L_4 \times \frac{P}{T} + L_7 \times \frac{P^2}{T} \tag{S7}$$

Eq. (S7) has six unknowns and we have experimental data of Log[$f$(O$_2$)] at $P$ = (0GPa, 20GPa, 40GPa) and $T$ = (1500K, 2000K), which gives us six data points for fitting, so a simple regression fit will yield the unknown G values. We find that for a fit with $P$ in GPa and $T$ in 1000K (or 1/$T$ in units of 1000/$T$, so at $T$=1000K we have 1/$T$ = 1, $T$=2000K we have 1/$T$=0.5, etc.) we get:

$$Log[f(O_2)](P, \frac{1}{T}) =$$
$$7.1 + 1.75 \times 10^{-2} \times P + (-24) \times \frac{1}{T} + 1.25 \times 10^{-4} \times P^2 + 0.57 \times \frac{P}{T} + (-1.5 \times 10^{-3}) \times \frac{P^2}{T}$$
(S8)

The data used for fitting and the fitted parameters and the are shown in Table S1 and the values used for the thermodynamic model are shown in Table S2:

| P \ T | Log[$fO_2$] data used for fitting | |
|---|---|---|
| | 1500 K | 2000 K |
| 0 GPa | -8.9 | -4.9 |
| 20 GPa | -1.3 | 0.9 |
| 40 GPa | 5.6 | 6.2 |

Table S1. The data of Log[$fO_2$] used for fitting our Log[$fO_2$] model.

| P \ T | 2000 K | 3000 K | 4000 K |
|---|---|---|---|
| 20 GPa | 0.9 | 3.1 | 4.2 |
| 30 GPa | 3.61 | 4.97 | 5.66 |
| 40 GPa | 6.2 | 6.8 | 7.1 |
| 50 GPa | 8.66 | 8.54 | 8.48 |
| 60 GPa | 11 | 10.2 | 9.8 |
| 70 GPa | 13.21 | 11.79 | 11.1 |
| 80 GPa | 15.3 | 13.3 | 12.3 |
| 90 GPa | 17.26 | 14.74 | 13.48 |
| 100 GPa | 19.1 | 16.1 | 14.6 |
| 110 GPa | 20.82 | 17.39 | 15.68 |
| 120 GPa | 22.4 | 18.6 | 16.7 |

Table S2. Log[$fO_2$] of Re-ReO$_2$ capsule in lower mantle $T$ and $P$ range.

After we have a model for the Log[$f(O_2)$] value of the Re-ReO$_2$ capsule, the next step is trying to find the relation between Re-ReO$_2$ capsule and other capsules. The $fO_2$ in the diamond capsule is about 4 orders of magnitude lower than the Re-ReO$_2$ at 25GPa, 2000K (Pownceby and O'Neil, 1994). We assume that within the pressure and temperature range of our thermodynamic model ($P$: 20~120GPa, $T$: 2000~4000K), the diamond capsule $f$O$_2$ is always 4 orders of magnitude lower than the Re-ReO$_2$ capsule $f$O$_2$. We also make a quick estimation of the effect of this assumption, if we vary the diamond capsule $f$O$_2$ by ±2 order of magnitude, at T=2000K, the reaction energy change of Eq. (1) is about ±0.19eV/Fe. The corresponding Fe$^{3+}$/ΣFe difference is about +0.14 and -0.07 for $T$=2000K, diamond capsule case. Qualitatively this variance doesn't change the fact that the Fe$^{3+}$ fraction is small (0~0.35 within the lower mantle pressure range) in diamond capsule at $T$=2000K. And we still have the fact that the Fe$^{3+}$/ΣFe in diamond capsule is lower than that in Re-ReO$_2$ capsule. Therefore in our thermodynamic model, we use $(Log[f(O_2)]|_{Re-ReO_2} - 4)$ as the $f$O$_2$ input for the diamond capsule and we believe that likely deviations from this assumption will not have a qualitative effect on our results.

These relationships provide at least approximate models for the $f$O$_2$ for all the necessary conditions for the thermodynamic modeling in this work.

## 2. Integrated model details

Here we describe how we combine the oxidation reaction model and chg. disp. reaction model together. The approach is the same as with the models treated separately. Consider the equilibration process by starting with one mole of $(Mg_{1-x}Fe_x)SiO_3$. When the system comes into equilibrium, $n_1$ mole of ferrous Mg-Pv is transferred to ferric Mg-Pv through the oxidation reaction and $n_2$ mole of ferrous Mg-Pv is transferred to ferric Mg-Pv through the chg. disp. reaction. Then the compact form for the whole system is:

$$\left(1+\frac{1}{2}n_1x+\frac{1}{3}n_2x\right)\left(Mg_{\frac{1-(1-n_1-n_2)x}{1+\frac{1}{2}n_1x+\frac{1}{3}n_2x}}Fe^{2+}_{\frac{(1-n_1-n_2)x}{1+\frac{1}{2}n_1x+\frac{1}{3}n_2x}}Fe^{3+}_{\frac{n_1x/2+n_2x/3}{1+\frac{1}{2}n_1x+\frac{1}{3}n_2x}}\right)\left(Si_{\frac{1}{1+\frac{1}{2}n_1x+\frac{1}{3}n_2x}}\overset{HS}{Fe^{3+}}_{\frac{y(n_1x/2+n_2x/3)}{1+\frac{1}{2}n_1x+\frac{1}{3}n_2x}}\overset{LS}{Fe^{3+}}_{\frac{(1-y)(n_1x/2+n_2x/3)}{1+\frac{1}{2}n_1x+\frac{1}{3}n_2x}}\right)O_3$$

$$+\frac{1}{4}(1-n_1)xO_2+(1-n_1-n_2)xMgO+\frac{n_2x}{3}Fe^0$$

(S9)

We take the total Fe content as $x = 1/8$ and $y$ is the B-site HS $Fe^{3+}$ ratio. As done for the oxidation reaction and chg. disp. reaction separately, we can use Eq. S9 to write the expression of total Gibbs energy $G(n_1,n_2,y;P,T)$. The expression of total Gibbs energy in the integrated model is:

$$G_{Total}=(1-n_1-n_2)H[(Mg_{1-x}Fe_x)SiO_3]+(\frac{n_1}{2}+\frac{n_2}{3})yH[(Mg_{1-x}Fe_x)(Si_{1-x}\overset{HS}{Fe_x})O_3]+(\frac{n_1}{2}+\frac{n_2}{3})(1-y)H[(Mg_{1-x_1}Fe_{x_1})(Si_{1-x_1}\overset{LS}{Fe_{x_1}})O_3]$$

$$+(1-n_1-n_2)H[MgO]+\frac{1}{4}(1-n_1)\mu(O_2)+\frac{xn_2}{3}H[Fe]-(1-n_1-n_2)xTS_{mag}(Fe^{2+},A)-(\frac{n_1}{2}+\frac{n_2}{3})xTS_{mag}(Fe^{3+},A)$$

$$-(\frac{n_1}{2}+\frac{n_2}{3})xyTS_{mag}(Fe^{3+\,HS},B)-(\frac{n_1}{2}+\frac{n_2}{3})x(1-y)TS_{mag}(Fe^{3+\,LS},B)-\frac{xn_2}{3}TS[Fe^0]-(1+\frac{n_1x}{2}+\frac{n_2x}{3})TS_{conf}(A)-(1+\frac{n_1x}{2}+\frac{n_2x}{3})TS_{conf}(B)$$

(S10)

where $G_{Total}$ is the total Gibbs energy for the system with states associated with both the oxidation and chg. disp. reactions. The definitions of the enthalpy terms and the magnetic entropy terms are consistent with those in the oxidation reaction model. The configurational entropy terms in sublattices A and B are:

$$S_{config}(A)=k_B[\frac{1-(1-n_1-n_2)x}{1+\frac{n_1}{2}x+\frac{n_2}{3}x}\ln(\frac{1-(1-n_1-n_2)x}{1+\frac{n_1}{2}x+\frac{n_2}{3}x})+\frac{(1-n_1-n_2)x}{1+\frac{n_1}{2}x+\frac{n_2}{3}x}\ln(\frac{(1-n_1-n_2)x}{1+\frac{n_1}{2}x+\frac{n_2}{3}x})$$

$$+\frac{(\frac{n_1}{2}+\frac{n_2}{3})x}{1+\frac{n_1}{2}x+\frac{n_2}{3}x}\ln(\frac{(\frac{n_1}{2}+\frac{n_2}{3})x}{1+\frac{n_1}{2}x+\frac{n_2}{3}x})]$$

$$S_{config}(B) = k_B[\frac{1}{1+\frac{n_1}{2}x+\frac{n_2}{3}x}\ln(\frac{1}{1+\frac{n_1}{2}x+\frac{n_2}{3}x}) + \frac{y(\frac{n_1}{2}x+\frac{n_2}{3}x)}{1+\frac{n_1}{2}x+\frac{n_2}{3}x}\ln(\frac{y(\frac{n_1}{2}x+\frac{n_2}{3}x)}{1+\frac{n_1}{2}x+\frac{n_2}{3}x}) +$$

$$\frac{(1-y)(\frac{n_1}{2}x+\frac{n_2}{3}x)}{1+\frac{n_1}{2}x+\frac{n_2}{3}x}\ln(\frac{(1-y)(\frac{n_1}{2}x+\frac{n_2}{3}x)}{1+\frac{n_1}{2}x+\frac{n_2}{3}x})]$$

(S11)

We can obtain values for $n_1$, $n_2$, and $y$ by simultaneously minimizing $G_{TotalAll}$ with respect to these variables. Note that $\partial G/\partial y$ is independent of $n_1$ and $n_2$, which is consistent with the fact that the fraction of HS vs. LS B-site $Fe^{3+}$ does not depend on the extent of the chg. disp. reaction.

## 3. DFT setup information and EOS comparison

This SI section contains further details of our density functional theory calculation approach. The endmembers of our system are ferrous Mg-Pv $(Mg_{1-x}Fe_x)SiO_3$ (where $x = 1/8$), ferric Mg-Pv $(Mg_{1-x_1}Fe_{x_1})(Si_{1-x_1}Fe_{x_1})O_3$ (where $x_1 = 1/17$ in the oxidation reaction model and $x_1 = 1/25$ in the chg. disp. reaction), pure Mg-Pv, MgO, and metallic Fe. As our model assumes that Fe can be treated as weakly interacting on a given sublattice we attempt to obtain a consistent set of energetics by estimating the energy for maximally isolated Fe on each sublattice. We therefore use a 120 atom supercell for all the Mg-Pv structures with at most just one Fe atom on each sublattice, which are $(Mg_{23}Fe_1)(Si_{24})O_{72}$ for ferrous Mg-Pv and $(Mg_{23}Fe_1)(Si_{23}Fe_1)O_{72}$ for ferric Mg-Pv. All the details of $k$ points and symmetry information are shown in **Table S3**. The choices of $k$ point mesh yield a convergence of total reaction energy better than 10 meV/Fe. Structural relaxation is set to converge to $10^{-3}$ eV in the total energy, yielding the average forces between atoms to be about 0.01 eV/Å.

|  | DFT set up information | | | |
|---|---|---|---|---|
|  | Number of atoms | k-points | space group | Calculated lattice constant (20 GPa) (Ang) axbxc |
| MgO | 2 | 5x5x5 | $Fm\bar{3}m$ | 2.04 (Mg-O bond) |
| $Mg_{24}Si_{24}O_{72}$ | 120 | 1x1x2 | Pnma | 9.31 x 14.49 x 6.73 |
| $(Mg_{23}Fe_1)Si_{24}O_{72}$ | 120 | 1x1x2 | Pnma | 9.31 x 14.50 x 6.74 |
| $(Mg_{23}Fe_1)(Si_{23}Fe_1)O_{72}$ HH | 120 | 1x1x2 | Pnma | 9.32 x 14.55 x 6.76 |
| $(Mg_{23}Fe_1)(Si_{23}Fe_1)O_{72}$ HL | 120 | 1x1x2 | Pnma | 9.31 x 14.52 x 6.75 |
| Metallic hcp Fe | 2 | 5x5x5 | $P6_3/mmc$ | a=2.62 c=4.4 (0GPa) |

**Table S3.** The ab initio parameters used for the compounds in our model. We use experimental EOS for the metallic hcp Fe combined with the DFT calculation at P = 0 GPa, so we show the calculated lattice constant of hcp Fe at 0 GPa here. "HH" means high spin in both A and B sites for ferric Mg-Pv and "HL" means high spin in A site and low spin in B site.

For the $Fe^{3+}$-$Fe^{3+}$ coupled ferric Mg-Pv, we use a $(Mg_{23}Fe_1)(Si_{23}Fe_1)O_{72}$ supercell with one Fe in the A site and the other Fe in the B site. The two Fe atoms are placed as neighbors to obtain a configuration with the lowest energy (Zhang and Oganov, 2006). We also have two different spin states for the ferric Mg-Pv. At relatively low pressures, the coupled $Fe^{3+}$-$Fe^{3+}$ is in the high-spin and high-spin (HS-HS) state. When the pressure increases, the ferric Mg-Pv will gradually change to a high-spin and low-spin (HS-LS) state. The HS-HS corresponds to the state with $5\mu_B$ and $5\mu_B$ of magnetic moment on the A site and B site respectively. The HS-LS is the state with $5\mu_B$ and $1\mu_B$ on the A site and B site.

To validate the accuracy of our ab-initio calculation of these high pressure phases, we compare the calculated equation of state (EOS) parameters ($V_0$, $K_0$, $K_0{'}$) of all the endmembers in our thermodynamic model with experiments. The comparison is shown in **Table. S4.** For $V_0$, the relative difference ($|V_{0,exp} - V_{0,sim}|/V_{0,exp}$) in all the cases is less than 0.5% and calculated values are typically smaller. The zero pressure bulk modulus relative difference ($|K_{0,exp} - K_{0,sim}|/K_{0,exp}$) in all the cases is less than 3.5%. We note that our DFT results are at 0 K, but the experimental results correspond to 300 K. So we need to consider the thermal effects on volume and bulk modulus. At room temperature and 1 bar condition, the thermal expansion coefficient, $\alpha = \frac{1}{V}(\frac{dV}{dT})_P$, and temperature derivative of bulk modulus $K$ with respect to pressure, $(\frac{dK}{dT})_P$, are: (1) for Mg-Pv, $\alpha = 2.5 \times 10^{-5}$/K (Tange et al., 2012), $(\frac{dK}{dT})_P = -0.017$ GPa/K (Fiquet et al., 2000). (2) for MgO, $\alpha = 3.56 \times 10^{-5}$/K and $(\frac{dK}{dT})_P = -0.015$ GPa/K (Mao et al., 2011a). These thermal expansions are not fit to low temperatures below the Debye temperature but should provide an upper bound for the scale of changes we are likely to see in these materials from 0 to 300 K. With temperature difference $\Delta T = 300$ K the thermal effect on volume is about 0.75% change for Mg-Pv and 1% change for MgO, and the thermal effect on bulk modulus is about 2% change for Mg-Pv, 3% change for MgO. So the temperature effect (from 0 to 300 K) is expected to be close to or within our simulation uncertainties. Based on the above discussion, our calculated EOS parameters are in good agreement with those from experiments, which overall gives us confidence in the validity of our DFT energetics.

|  | Simulation | | | | Experiment | | | |
| --- | --- | --- | --- | --- | --- | --- | --- | --- |
|  | $V_0$ (Ang^3) /f.u. | $K_0$ (GPa) | $K_0'$ | Fe % /site | $V_0$ (Ang^3)/f.u. | $K_0$ (GPa) | $K_0'$ | Fe %/site |
| MgO | 18.62 | 166.7 | 4.15 | N/A | 18.67 [a] | 160 [b] | 4 [b] | N/A |
| $Mg_{24}Si_{24}O_{72}$ | 40.5 | 256.14 | 4.02 | N/A | 40.58 [c] | 261 [c] | 4 [c] | N/A |
| $(Mg_{23}Fe_1)Si_{24}O_{72}$ | 40.58 | 255.04 | 4.06 | 4.2% | 40.65 [c] | 260.4 [c] | 4 [c] | 4.2% |
| $(Mg_{23}Fe_1)(Si_{23}Fe_1)O_{72}$ HH | 40.93 | 252.02 | 4.04 | 4.2% | 41.00 [d] | 249.23 [d] | 4 [d] | 4.2% |
| $(Mg_{23}Fe_1)(Si_{23}Fe_1)O_{72}$ HL | 40.59 | 271.2 | 3.69 | 4.2% | 40.42 [d] | 282.08 [d] | 4 [d] | 4.2% |
| Metallic hcp Fe | — | — | — | N/A | 11.19 [e] | 156 [e] | 5.81 [e] | N/A |

[a] (Jacobsen et al., 2002)

[b] (Speziale et al., 2001)

[c] (Lundin et al., 2008)

[d] (Catalli et al., 2010)

[e] (Dubrovinsky et al., 2000)

**Table S4.** Fitted 3$^{rd}$-order Birch-Murnaghan EOS parameters of all the compounds in the ab-inito based thermodynamic model (fit over a pressure range of -10GPa to 120GPa), compared to the results of experiments. "/f.u." means per formula unit, and for all the Mg-Pv endmembers one f.u. corresponds to one A, one B, and 3 O sites (a 5-site $ABO_3$ unit cell). "HH" means high spin in both A and B sites for ferric Mg-Pv and "HL" means high spin in A site and low spin in B site. As we take experimental EOS parameters for the metallic Fe in our thermodynamic model, we don't show the simulated values for these parameters of metallic Fe.

For the metallic Fe, experiments show that the stable phase in the lower mantle pressure range is hcp in a non-magnetic state (Mathon et al., 2004). However, there is presently a disagreement about the electronic structure of Fe between simulations and experiments (Sha and Cohen, 2010). Furthermore, the EOS of hcp Fe at high pressure by simulation does not match well with experiments. Therefore, in this work, we use only the ab initio energy for hcp metallic Fe at 0 GPa and otherwise use the experimental Birch-Murnaghan third order equation of state parameters for metallic Fe (Dubrovinsky et al., 2000), which are given in **Table S4**. For the ab-initio energy of hcp metallic Fe at 0 GPa, we use the lowest energy state (ground state) value, which is the 0 GPa energy of ferromagnetic hcp metallic Fe. With our ab initio energy and this EOS we can calculate the enthalpy of Fe at any given pressure.

We have chosen not to use the more commonly applied DFT+$U$ (Anisimov et al., 1991) method for treating these particularly correlated transition metal oxide systems. DFT+$U$ has been used for a wide-range of studies of Fe in geophysical studies, including in both (Mg,Fe)O ferropericlase (Persson et al., 2006; Tsuchiya et al., 2006; Wentzcovitch et al., 2009) and Mg-Pv (Bengtson et al., 2008; Hsu et al., 2011, 2010; Metsue and Tsuchiya, 2012). However, while $U$ can be determined by self-consistent ab initio methods, such calculations for Fe in ferropericlase and Mg-Pv have shown wide variation in $U$ values for Fe depending on site, valence, spin state and pressure (Hsu et al., 2011, 2010; Tsuchiya et al., 2006), with values ranging from 3 to 5 eV for the different possible Fe states in just Mg-Pv (Hsu et al., 2011, 2010). Given the many environments and the wide range of pressures required to treat Fe in this work, including metallic $Fe^0$, the hybrid HSE06 functional appeared to provide a simpler approach, without a large set of $U$ values to establish. To assure that our results were not completely different with the two methods we studied the chg. disp. reaction (Eq. (9)) with the GGA+$U$ method using self-consistent $U$ values for A and B sites taken from Hsu et al. (2011, 2010). These results found a $\Delta H$ = 400 meV/Fe, comparable to the value of 500 meV/Fe for the HSE results shown in this work. We also note that our spin transition pressure, discussed in section 3.2, matches closely with previous LDA+$U$ calculations. Therefore, we believe that similar results as those presented here could also be obtained by careful application of varied self-consistent $U$'s, although that method was not pursued here.

## 4. Impact of Fe partitioning

In this section we consider the potential impact of partitioning of Fe between Mg-Pv and ferropericlase (Fp) on our predictions for $Fe^{3+}/\Sigma Fe$ in Mg-Pv. We will split the problem into two parts: (a) at high $fO_2$, only oxidation reaction occurs in the system; (b) at low $fO_2$, we will study what is the $fO_2$ level when chg. disp. reaction suppresses oxidation reaction.

### 4.1 At high $fO_2$

As the oxidation reaction is the dominant reaction at high $fO_2$ condition (by which we mean $fO_2$ higher than the $fO_2^t$ identified in section 3.1, we don't need to consider the chg. disp. reaction. If some amount of Fe partitions from Mg-Pv to Fp, two effects may change the $Fe^{3+}/\Sigma Fe$ calculated from Mg-Pv coexisting with pure MgO model (Eq. 1 and Eq. 9).

First, the effective chemical potential of MgO is lowered if we have some Fe in the Mg site. If we assume that the FeO and MgO mix as an ideal solution in Fp, the effective chemical potential of MgO in $(Mg_{1-x}Fe_x)O$ can be estimated as:

$$\mu^{eff}(MgO) = \mu^0(MgO) + k_B T \ln(1-x) \tag{S12}$$

In the lower mantle, the Fe in Fp is about x=0.2 (Hirose, 2002). Here we will show $Fe^{3+}/\Sigma Fe$ in Mg-Pv when Fe in Fp is $x = 0.1$, 0.2, and 0.3 to assess the influence of MgO chemical potential changes from Fe partitioning into MgO. The results are shown in Fig. S1, where the predictions have been made with Eq. 1 with the MgO chemical potential modified by Eq. S12. We can see that there is no significant change of $Fe^{3+}/\Sigma Fe$ in Mg-Pv. Even when the Fe in Fp is $x = 0.3$, the largest decrease of $Fe^{3+}/\Sigma Fe$ in Mg-Pv is less than 0.08. Therefore our reactions (Eq. 1) with pure MgO as the reactants give qualitatively and even semiquantitatively the correct values for what would be obtained with equilibrium with (Mg,Fe)O.

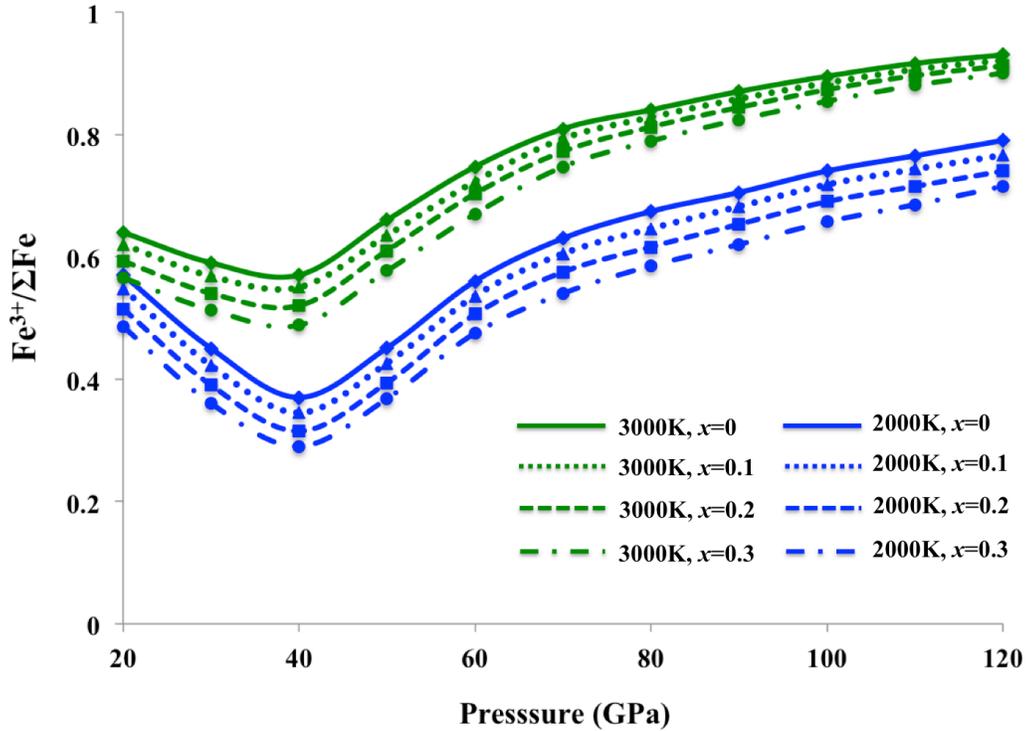

**Fig. S1** $Fe^{3+}/\Sigma Fe$ in Mg-Pv at Re-ReO$_2$ capsule $fO_2$ condition when Fe% in Fp is 10, 20 and 30. The blue curves correspond to T=2000K, the green curve correspond to T=3000K. The solid lines are the original $Fe^{3+}/\Sigma Fe$ curves when Fe content in $(Mg_{1-x}Fe_x)O$ is $x = 0$. The dotted lines, dashed lines and dotted-dashed lines correspond to Fe content in Fp of $x = 0.1$, 0.2, and 0.3, respectively.

A second effect due to the Fe partitioning between Mg-Pv and Fp is that the total Fe content in Mg-Pv might vary. Here we show that the variation of $Fe^{3+}/\Sigma Fe$ in Mg-Pv arising from a change of Fe content in Mg-Pv is quite small and can be neglected. Table. S5 shows the result of $Fe^{3+}/\Sigma Fe$ at different Fe% in Mg-Pv and different temperature and pressure conditions. These results indicate that although the Fe partitioning between Mg-Pv and Fp may alter the total Fe content in Mg-Pv, the $Fe^{3+}/\Sigma Fe$ ratio predicted from our thermodynamic model changes by less than 10% over a very large range of Mg-Pv Fe content. These results show that our results for $Fe^{3+}/\Sigma Fe$ ratio in Mg-Pv are quite general for different Fe in Mg-Pv compositions and, in particular, support that we can neglect effects of Fe partitioning in the model.

| Total Fe% in Mg-Pv | $Fe^{3+}/\Sigma Fe$ in Mg-Pv | | | |
|---|---|---|---|---|
| | T=2000K | | T=3000K | |
| | 40GPa | 100GPa | 40GPa | 100GPa |
| 1% | 0.35 | 0.697 | 0.558 | 0.879 |
| 5% | 0.362 | 0.725 | 0.565 | 0.891 |
| 12.5% | 0.37 | 0.74 | 0.575 | 0.895 |
| 20% | 0.39 | 0.75 | 0.591 | 0.903 |

**Table. S5** The dependence of $Fe^{3+}/\Sigma Fe$ on the total Fe content in Mg-Pv at different pressure and temperature conditions. 40GPa represents relatively low pressures region. 100GPa represents relatively high pressures region. The result of total Fe% in Mg-Pv=12.5% corresponds to the prediction of our model in the main text.

### 4.2 At low $fO_2$

When the $fO_2$ goes lower than the $fO_2^t$ value discussed in section 3.1 the influence on the $Fe^{3+}/\Sigma Fe$ ratio in Mg-Pv due to Fe partitioning is similar to that shown above. Now we only need to consider chg. disp. reaction (Eq. 9). We still have two effects that may change the $Fe^{3+}/\Sigma Fe$ ratio in Mg-Pv when we consider the Fe partitioning. First, the effective chemical potential of MgO is lowered, which is shown in Eq. S12. Second, the Fe content in Mg-Pv may change. The impact of these two effects on the chg. disp. reaction should be similar to that on the oxidation reaction, which has already been discussed in SI section 4.1. The general conclusion is that the Fe partitioning also doesn't significant change the $Fe^{3+}/\Sigma Fe$ in Mg-Pv contributed from chg. disp. reaction. To consider the $fO_2$ transition point between chg. disp. reaction and oxidation reaction, please refer to the discussion in the section 3.1 and the details shown in SI section 5.

## 5. Ideal solution model of $Fe^0$-ferropericlase equilibrium $fO_2$

The possible range of the $Fe^0$-ferropericlase (Fp) equilibrium $fO_2$ (shown in the grey box in Fig.1) is derived from experimental data and an ideal solution model. We model Fe-FeO capsule relative to Re-ReO$_2$ capsule following the previous experimental evidence suggesting that the $fO_2$ of Fe-FeO capsule is about 6 order of magnitude lower than that of Re-ReO$_2$ capsule (Frost et al., 2004). We then further modify the $\mu(Fe^{2+})$ to correct the fact that the FeO is dissolved in MgO to form the Fp. Based on the reaction $2Fe^0 + O_2 \rightarrow 2Fe^{2+}O$ and the expression of $\mu(Fe^{2+})$ in $(Mg_{1-x}Fe_x)O$: $\mu(Fe^{2+}) = \mu(Fe^{2+})_{FeO} + kT\ln x$ (ideal solution model) we can easily calculate the $fO_2$ shift relative to Fe-FeO level at a certain Fe concentration in ferropericlase.

First, we define the equilibrium $fO_2$ of Fe-FeO system as $fO_2$[Fe-FeO]. Then we have the equation:

$$2\mu(Fe^0) + \mu_{O2} + kT\ln fO_2[Fe\text{-}FeO] = 2\mu(FeO) \tag{S13}$$

If the system is at $P$=100GPa, $T$=2000K which is consistent with the condition in Fig. 1, $\mu(Fe^0)$ is the chemical potential of metallic Fe at (100GPa, 2000K), $\mu(FeO)$ is the chemical potential of FeO at (100GPa, 2000K), $\mu_{O2}$ is the reference chemical potential of O$_2$ when partial pressure of O$_2$ is 1atm at $T$=2000K.

Then we define the equilibrium $fO_2$ of Fe- $(Mg_{1-x}Fe_x)O$ system as $fO_2[(Mg_{1-x}Fe_x)O]$. Then we have the equation:

$$2\mu(Fe^0) + \mu_{O2} + kT\ln fO_2[(Mg_{1-x}Fe_x)O] = 2(\mu(FeO) + kT\ln x) \tag{S14}$$

Then we do (Eq. S14) – (Eq. S13), it yields the following expression:

$$\text{Log}[fO_2[(Mg_{1-x}Fe_x)O]] = \text{Log}[fO_2[Fe\text{-}FeO]] + 2\text{Log}x = fO_2[\text{Re-ReO}_2\text{-}6] + 2\text{Log}x \tag{S15}$$

With Eq. S15, we are able to create the grey box shown in Fig. 1 which tells us the range of the equilibrium $fO_2$ of $(Mg_{1-x}Fe_x)O$ where $x$ is from 0.05 to 0.4.

Erratum for the Paper "*Origin of $Fe^{3+}$ in Fe-containing, Al-free Mantle Silicate Perovskite*"

Shenzhen Xu[1], Sang-Heon Shim[3], Dane Morgan[1,2]

[1]*Materials Science Program*

[2]*Department of Materials Science and Engineering*

*University of Wisconsin – Madison, Madison, WI*

[3]*School of Earth and Space Exploration*

*Arizona State University, Tempe, AZ*

Correpsonding author (Dane Morgan): Telephone: +1-608-265-5879; Fax: +1-608-262-8353; Email: ddmorgan@wisc.edu

In the main text section 2.1.1 the term $H^0_{vib}(O^{2-}_{solid})$ in the equation for $G_{vib}(O^{2-}_{solid}) - H^0_{vib}(O^{2-}_{solid})$ was miscalculated. The incorrect value was 0.63eV and the corrected value is 0.095eV. As discussed below, this correction demonstrated that there was some discrepancy in the DFT oxidation energies, so we now add another correction term with an increase $\mu(O_2)$ in main text eq. (4) by 0.4eV/$O_2$. These corrections change the reaction energy of Eq. (1), and have the effect of stabilizing oxygen gas and reducing the amount of $Fe^{3+}$ created by oxidation. This change does not significantly impact the curve shapes, the qualitative conclusions, or the discussions, except regarding Fig. 1, which we detail below. Unfortunately, many of the specific values shown in the figures and mentioned in the text related to the oxidation reaction in Eq. (1) are somewhat changed, so below we give revised figures and specific corrections for regions of text or values which need to be updated. Figs. 1, 2, 4(a), and 5 in the main text are changed to the new Figs E1, E2, E4(a), and E5, respectively.

The changes to the text are: **Pg 323/lc(**left column**) ln(**line**) 11**, remove "We don't use any data from FeO or ferropericlase in constructing our model, so being consistent with $Fe^0$/ferropericlase equilibrium thermodynamics is an important test of the model." **Pg 323/lc ln33**, remove "The ability to define this range consistently for both Mg-Pv and Fp, despite the model being developed without any explicit *ab initio* calculations on the Fp system, supports the accuracy of our thermodynamic model." **Pg 323/lc ln 25**, change "11.6" to "12.8". **Pg 323/lc ln 33**, change "11.6" to "12.8". **Pg 323/rc(**right colunm**) ln 22**, change "0.5" to "0.3". **Pg 323/rc ln 22**, change "0.08" to "0.05". **Pg 323/rc ln 24**, change "0.7" to "0.6". **Pg 323/rc ln 24**, change "0.2" to "0.13". **Pg 326/lc ln 9**, change "0.4-0.5" to "0.3-0.35". **Pg 326/lc ln 9**, change "0.08-0.1" to "0.05". **Pg 326/lc 2$^{nd}$ ln from the bottom**, change "0.05" to "0.03". **Pg 326/rc ln 1**, change "0.17" to "0.035". **Pg 326/rc ln 1**, change "0.1" to "0.045". **Pg 326/rc ln 2**, change "0.89" to "0.19".**Pg 326/rc ln 2**, change "0.52" to "0.24".

Our correction to $H^0_{vib}(O^{2-}_{solid})$ changes the predicted transition oxygen fugacity, $fO_2^t$, which is where the charge disproportionation (chg. disp.) reaction ceases to occur and the oxidation reaction starts to occur. Specifically, in Fig. 1 (main text) $fO_2^t$ changes from $\log(fO_2^t)$ = $\log(fO_2[Re-ReO])$ - 6.9) to $\log(fO_2^t)$ = $\log(fO_2[Re-ReO])$ - 4.7. Because the bridgmanite (Pv) is in equilibrium with ferropericlase (Fp) (Mg,Fe)O the Fe from the chg. disp. reaction can only exist for oxygen fugacity less than $fO_2[Fe-(Mg,Fe)O]$. We show below that $fO_2[Fe-(Mg,Fe)O]$ = $\log(fO_2[Re-ReO_2])$ – 5.7. This yields a final change in value from the incorrect value $\log(fO_2^t)$ = $\log(fO_2[Re-ReO])$ - 6.9) to the correct value $\log(fO_2^t)=\log(fO_2[Re-ReO_2])$ – 5.7. So in our corrected Fig. E1, we can see the $\log(fO_2^t)$ is at $\log(fO_2[Re-ReO_2])$ – 5.7. The determination of $\log(fO_2[Fe-(Mg,Fe)O])$ is as follows. We first consider relative difference between $fO_2[Fe-FeO]$ and $fO_2[Re-ReO_2]$ at $T$=2000K, $P$=20GPa-100GPa (lower mantle relevant conditions) from experiments (Campbell et al., 2007; Campbell et al., 2009). The experimental results show that $\log(fO_2[Fe-FeO])$ is -4.5 to -5 with respect to $\log(fO_2[Re-ReO_2])$. We take an average which is -4.75≈-4.8 to represent the $\log fO_2$ difference between Fe-FeO and Re-ReO$_2$. Now we consider the different between $\log(fO_2[Fe-FeO])$ and $\log(fO_2[Fe-(Mg,Fe)O])$. In our model the Fe content in Pv is 0.125 per formula unit of MgSiO$_3$, which is $(Mg_{0.875}Fe_{0.125})SiO_3$. Based on previous experiments, the Fe partitioning coefficient $K_D$ between Pv and Fp is 0.2-0.3 in Al-free condition (Irifune, 2010). If we take an average, then we can assume $K_D$≈0.25 in Al-free condition. Then based on the definition of $K_D$ ($K_D^{Pv\text{-}Fp}$=(Fe/Mg)$_{Pv}$/(Fe/Mg)$_{Fp}$), we can find that (Mg$_{0.875}$Fe$_{0.125}$)SiO$_3$ should be in equilibrium with (Mg$_{0.64}$Fe$_{0.36}$)O. Under an ideal solution approximation (SI section 5), log($f$O$_2$[(Mg$_{0.64}$Fe$_{0.36}$)O]) is calculated to be -0.9 with respect to log($f$O$_2$[Fe-FeO]). Therefore based on above arguments, we have log($f$O$_2$[Fe-(Mg,Fe)O])= log($f$O$_2$[(Mg$_{0.875}$Fe$_{0.125}$)SiO$_3$]) = log($f$O$_2$[(Mg$_{0.64}$Fe$_{0.36}$)O]) = log($f$O$_2$[Fe-FeO]) – 0.9 = log($f$O$_2$[Re-ReO$_2$]) – 5.7 .

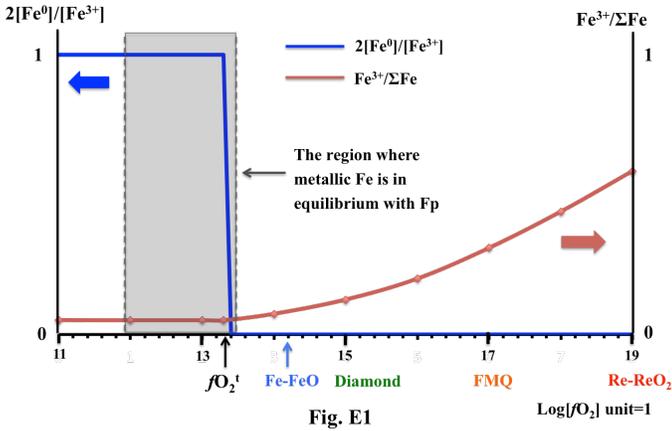

Fig. E1

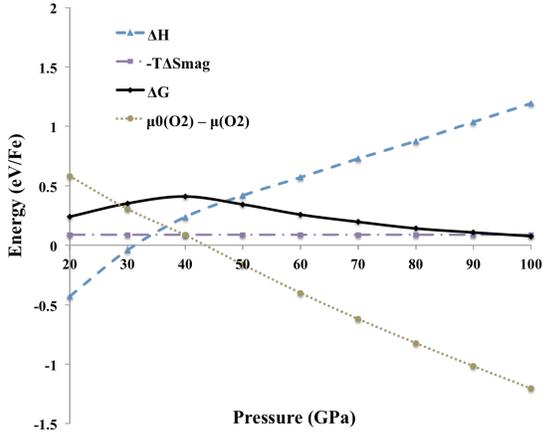

Fig. E4(a)

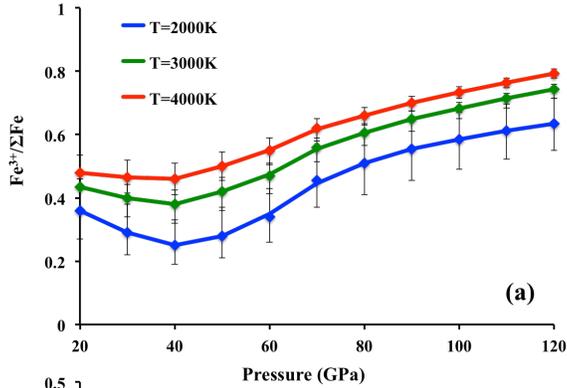

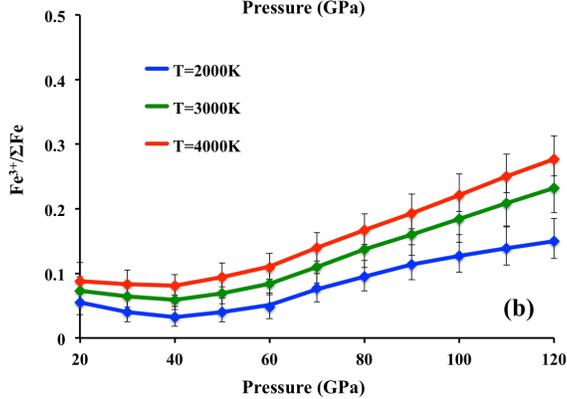

Fig. E2

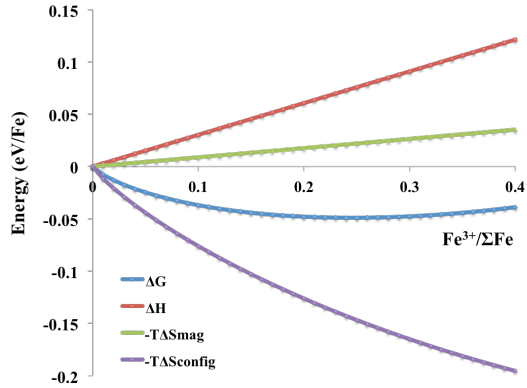

Fig. E5